\documentclass[reprint,amsmath,amssymb,aps,superscriptaddress,longbibliography]{revtex4-2}
\usepackage{xcolor}
\usepackage{graphicx}
\usepackage{dcolumn}
\usepackage{bm}
\usepackage{hyperref}

\hypersetup{%
   colorlinks = true,
   linkcolor=blue,
   citecolor=blue,
   urlcolor=blue,
}

\usepackage{soul}
\usepackage{units}

\begin{document}

\preprint{APS/123-QED}

\title{
Ultracold LiCr: a new pathway to quantum gases of  paramagnetic polar molecules
}
\author{S. Finelli}
\affiliation{Dipartimento di Fisica e Astronomia, Universit\`{a} di Firenze, 50019 Sesto Fiorentino, Italy}
\affiliation{Istituto Nazionale di Ottica del Consiglio Nazionale delle Ricerche (CNR-INO), 50019 Sesto Fiorentino, Italy}
\affiliation{\mbox{European Laboratory for Non-Linear Spectroscopy (LENS), Universit\`{a} di Firenze, 50019 Sesto Fiorentino, Italy}}
\author{A. Ciamei}
\email{ciamei@lens.unifi.it}
\affiliation{Istituto Nazionale di Ottica del Consiglio Nazionale delle Ricerche (CNR-INO), 50019 Sesto Fiorentino, Italy}
\affiliation{\mbox{European Laboratory for Non-Linear Spectroscopy (LENS), Universit\`{a} di Firenze, 50019 Sesto Fiorentino, Italy}}
\author{B. Restivo}
\affiliation{Istituto Nazionale di Ottica del Consiglio Nazionale delle Ricerche (CNR-INO), 50019 Sesto Fiorentino, Italy}
\affiliation{\mbox{European Laboratory for Non-Linear Spectroscopy (LENS), Universit\`{a} di Firenze, 50019 Sesto Fiorentino, Italy}}
\author{M. Schemmer}
\affiliation{Istituto Nazionale di Ottica del Consiglio Nazionale delle Ricerche (CNR-INO), 50019 Sesto Fiorentino, Italy}
\affiliation{\mbox{European Laboratory for Non-Linear Spectroscopy (LENS), Universit\`{a} di Firenze, 50019 Sesto Fiorentino, Italy}}
\author{A. Cosco}
\affiliation{Dipartimento di Fisica e Astronomia, Universit\`{a} di Firenze, 50019 Sesto Fiorentino, Italy}
\author{M. Inguscio}
\affiliation{\mbox{European Laboratory for Non-Linear Spectroscopy (LENS), Universit\`{a} di Firenze, 50019 Sesto Fiorentino, Italy}}
\affiliation{Department of Engineering, Campus Bio-Medico University of Rome, 00128 Rome, Italy}
\author{A.~Trenkwalder}
\affiliation{Istituto Nazionale di Ottica del Consiglio Nazionale delle Ricerche (CNR-INO), 50019 Sesto Fiorentino, Italy}
\affiliation{\mbox{European Laboratory for Non-Linear Spectroscopy (LENS), Universit\`{a} di Firenze, 50019 Sesto Fiorentino, Italy}}
\author{K. Zaremba-Kopczyk}
\affiliation{Faculty of Physics, University of Warsaw, Pasteura 5, 02-093 Warsaw, Poland}
\author{M. Gronowski}
\affiliation{Faculty of Physics, University of Warsaw, Pasteura 5, 02-093 Warsaw, Poland}
\author{M. Tomza}
\email{michal.tomza@fuw.edu.pl}
\affiliation{Faculty of Physics, University of Warsaw, Pasteura 5, 02-093 Warsaw, Poland}
\author{M. Zaccanti}
\affiliation{Istituto Nazionale di Ottica del Consiglio Nazionale delle Ricerche (CNR-INO), 50019 Sesto Fiorentino, Italy}
\affiliation{\mbox{European Laboratory for Non-Linear Spectroscopy (LENS), Universit\`{a} di Firenze, 50019 Sesto Fiorentino, Italy}}

\begin{abstract}
Quantum gases of doubly-polar molecules represent appealing frameworks for a variety of cross-disciplinary applications, encompassing quantum simulation and computation, controlled quantum chemistry and precision measurements. Through a joint experimental and theoretical study, here we explore a novel class of ultracold paramagnetic polar molecules combining lithium alkali and chromium transition metal elements. 
Focusing on the specific bosonic isotopologue $^{6}$Li$^{53}$Cr, leveraging on the Fermi statistics of the parent atomic mixture and on suitable Feshbach resonances recently discovered, we produce up to $50\times10^3$  ultracold LiCr molecules at peak phase-space densities exceeding 0.1, prepared within the least-bound rotationless level of the LiCr electronic \textit{sextet} ground state $X^6\Sigma^+$. 
We thoroughly characterize the  molecular gas, demonstrating  the paramagnetic nature of LiCr dimers and the precise control of their  quantum state. We investigate their stability against inelastic processes and identify a parameter region where pure LiCr samples exhibit lifetimes exceeding 0.2\,s. Parallel to this, we employ state-of-the-art quantum-chemical calculations to predict the properties of LiCr ground and excited electronic states. We identify efficient paths to coherently transfer weakly-bound LiCr dimers to their absolute ground state, to deliver ultracold gases of doubly-polar molecules with significant electric (3.3\,D) and magnetic ($5\,\mu_\text{B}$) dipole moments.
\end{abstract}

\pacs{Valid PACS appear here}
\maketitle

\section{Introduction}
\label{Sec:Intro}
Ultracold gases of ground-state molecules with permanent electric dipole moments offer unprecedented opportunities to investigate quantum chemistry and many-body physics \cite{Carr_2009,doi:10.1021/cr2003568,doi:10.1126/science.aam6299}. Recently, doubly-polar molecules, possessing both an electric and a magnetic dipole moment, have attracted great attention, as they could offer even richer prospects: While the electric dipole gives rise to the long-range anisotropic interactions, the magnetic dipole arising from a non-zero electronic spin provides an additional degree of tunability in the system. A high phase-space density gas of such molecules will open up new venues in the context of quantum simulation \cite{Micheli2006ToolboxPolMol,Pérez-Ríos_2010,Yao2018,CornishRev} and computation \cite{PhysRevA.101.062308,10.1063/1.4942928}, as well as quantum-controlled chemistry \cite{doi:10.1126/science.abl7257}. This is exemplified by the pioneering work on excited-state NaLi dimers \cite{PhysRevLett.119.143001}, which, despite featuring weak electric and magnetic dipole moments, provided new insights into ultracold reactive collisions \cite{doi:10.1126/science.abl7257,Park2023}.

Despite tremendous progress in the field \cite{Valtolina2023}, realization of degenerate gases of doubly-polar ground-state molecules remains an unsurpassed challenge. While direct laser cooling schemes have been successfully applied to doubly-polar radicals \cite{FITCH2021157}, the only experimental realizations of quantum degenerate molecular gases \cite{Valtolina2020,Duda2023,Bigagli2023} exploit ultracold atomic mixtures, where atom pairs are first associated into weakly-bound molecules across a Feshbach resonance (FR) \cite{RevModPhys.82.1225} and later transferred to the the absolute molecular ground-state via Stimulated Raman Adiabatic Passage (STIRAP). However, this two-step method has only been demonstrated on bi-alkali systems, whose ground state has zero electronic spin, thus negligible magnetic moment. Binding alkali species with closed-shell atoms, which would endow ground-state molecules with the additional spin and magnetic moment, are currently under investigation, following the experimental discovery of FRs in alkali--alkaline-earth systems \cite{RbSrFFR, PhysRevX.10.031037,PhysRevResearch.4.043072}. Yet, the extremely narrow character of available FRs, despite impressive technical effort \cite{10.1063/5.0143825}, has so far hindered the first association step. Another proposed pathway to realize paramagnetic ground-state molecules relies on binding highly-magnetic lanthanides with either alkali or closed-shell elements \cite{PhysRevX.10.041005}, and association of ultracold KDy Feshbach dimers has been indeed recently demonstrated \cite{PhysRevResearch.5.033117}. Nonetheless, STIRAP transfer of such molecules to their  ground state appears a formidable challenge, owing to the lack of spectroscopic data combined with the complex electronic configuration of these elements, which makes \textit{ab initio} methods poorly reliable, if not unapplicable.

\begin{figure*}[ht]
\centering
  \includegraphics[width=\textwidth]
  {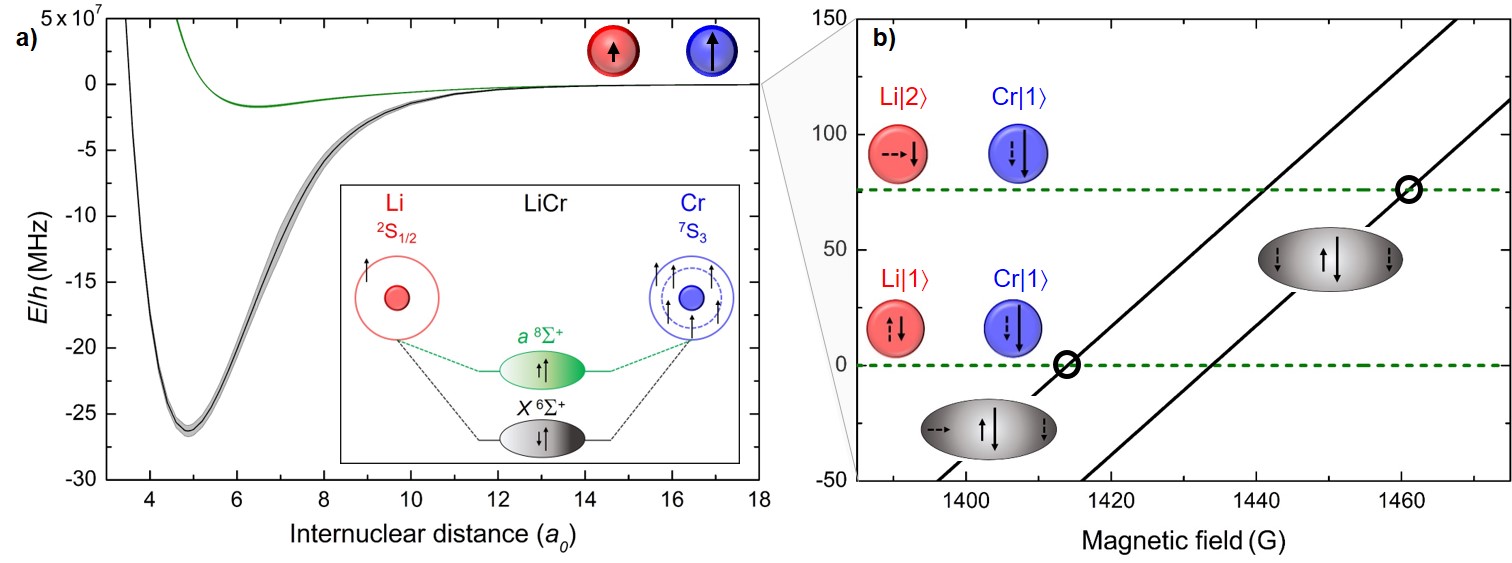}
  \caption{Doubly-polar LiCr dimers. (a) \textit{Ab initio} potential energy curves for the $X^6\Sigma^+$ ground-state and the $a^8\Sigma^+$ first excited state in Hund's case (a) representation. The inset shows a pictorial representation of the valence electrons for Li and Cr and the resulting sextet and octet molecules with non-zero electronic spin and polarized charge density. (b) Energy of the atom pairs in the Li$|1\rangle$-Cr$|1\rangle$ and Li$|2\rangle$-Cr$|1\rangle$ channels, of pure octet character (dashed green lines), and of the molecular sextet levels (solid black lines) inducing the FRs used in this work (circled). Atomic Zeeman levels are labeled in order of increasing energy. Solid and dashed arrows indicate electronic and nuclear spins, respectively.} 
  \label{fgr:OverviewMerged}
\end{figure*}

Here, through a joint experimental and theoretical study, we explore an alternative route towards ultracold doubly-polar molecules by binding an alkali atom (A) with transition-metal chromium (Cr). The interaction between A ($^2$S$_{1/2}$) and Cr ($^7$S$_3$) has only weak magnetic anisotropy and leads to two spin multiplicities, similarly to standard bi-alkali systems. While the latter feature a singlet $X^1\Sigma^+$ ground- and a triplet $a^3\Sigma^+$ first-excited-state \cite{Chin2010}, ACr systems exhibit a sextet $X^6\Sigma^+$ and an octet $a^8\Sigma^+$ symmetry, see  Fig.~\ref{fgr:OverviewMerged}(a) for the LiCr case. In its ground state ACr is thus expected to combine a high electronic spin with a strong $\textrm{A}^{\delta+}\textrm{Cr}^{\delta-}$ dipolar character \cite{VANZEE1985524}. Most importantly, the existence of two multiplicities, combined with the isotropic van der Waals interactions, makes the creation of ACr experimentally feasible with current technology. On the one hand, it leads to non-chaotic, bialkali-like FR spectra, well-suited for molecule association. On the other, it allows for cutting-edge molecular structure calculations with predictive power over ground and lowest-lying electronic states.
The ACr systems thus emerge as a unique playground that allows both to produce loosely-bound paramagnetic Feshbach dimers and to subsequently transfer them to their absolute ground state, leveraging on the ultracold atom toolbox developed for bi-alkalis. This provides a bridge between the well-known bi-alkali dimers and more complex molecular species. 

While early attempts focused on the rubidium-chromium mixture \cite{PhysRevA.81.052706,Hensler2004}, we here explore the lithium-chromium system, and in particular the $^6$Li$^{53}$Cr isotopologue, that combines the sole two stable fermionic isotopes of Li and Cr into bosonic dimers. This specific choice is motivated by the fact that Li-Cr Fermi mixtures have been recently realized \cite{PhysRevA.106.053318}, and that a suitable set of interspecies FRs has been identified \cite{PhysRevLett.129.093402}, see Fig.~\ref{fgr:OverviewMerged}(b). We experimentally demonstrate the efficient production of ultracold samples of up to $50 \times 10^3$ paramagnetic LiCr Feshbach dimers, at phase-space densities (PSD) exceeding 0.1 and temperatures of about 200\,nK. We exploit two inter-species $s$-wave FRs, see Fig.~\ref{fgr:OverviewMerged}(b), to access, with full internal quantum-state control, the least-bound rotationless vibrational level of the LiCr sextet ground state. This is confirmed by accurate measurements of the dimer magnetic moment and of the associated open-channel fraction, both found in remarkable agreement with theoretical expectations. Additionally, investigation of the collisional stability of the LiCr dimers as a function of the magnetic-field detuning from the resonance pole allows us to identify a parameter region in which the molecular Bose gas can exhibit lifetimes exceeding $0.2$ s, thanks to Pauli suppression of inelastic losses enabled by the fermionic nature of our mixture constituents \cite{Petrov2004, PhysRevA.94.062706}. Parallel to this, state-of-the-art \textit{ab initio} calculations, that we successfully benchmark against experimental data, are employed to quantify the doubly-polar nature of LiCr, and to identify optimal pathways for the coherent transfer of Feshbach dimers to the absolute ground state.

The paper is organized as follows. In Sec.~\ref{Sec:FermiMixAndProcedures} we summarize the experimental strategies to produce ultracold LiCr molecules and the properties of the parent atomic mixture under resonantly-interacting conditions. In Sec.~\ref{Sec:MoleculeFormation} we investigate the magneto-association process, including two-body adiabaticity, atom-to-molecule conversion efficiency, and the creation of a high phase-space density gas of LiCr. In Sec.~\ref{Sec:MoleculeProperties} we study single-particle properties of the newly formed Feshbach dimers, including their magnetic dipole moment and their binding energy, which we measure via a novel optical method. In Sec.~\ref{Sec:MoleculeLifetime} we identify and characterize the different loss mechanisms that limit the molecule lifetime and we show that under optimal conditions pure LiCr samples exhibit lifetimes exceeding 0.2\,s. In Sec.~\ref{Sec:Theory}, we present high-precision \textit{ab initio} potential energy curves (PECs) for LiCr, we demonstrate their predictive power on the atomic scattering properties, and then exploit them to unveil the doubly-polar character of their absolute ground-state. In Sec.~\ref{Sec:TheorySTIRAP}, we employ our model to explore electronically-excited states of LiCr, and identify an efficient STIRAP path connecting the last vibrational sextet level to the absolute ground state. Finally, in Sec.~\ref{Sec:Conclusions} we provide a conclusion and an outlook on future research directions.

\section{Resonantly interacting L\MakeLowercase{i}-C\MakeLowercase{r} Fermi mixtures and experimental procedures}
\label{Sec:FermiMixAndProcedures}
\begin{figure}[ht!]
    \centering
    \includegraphics[width=\columnwidth]{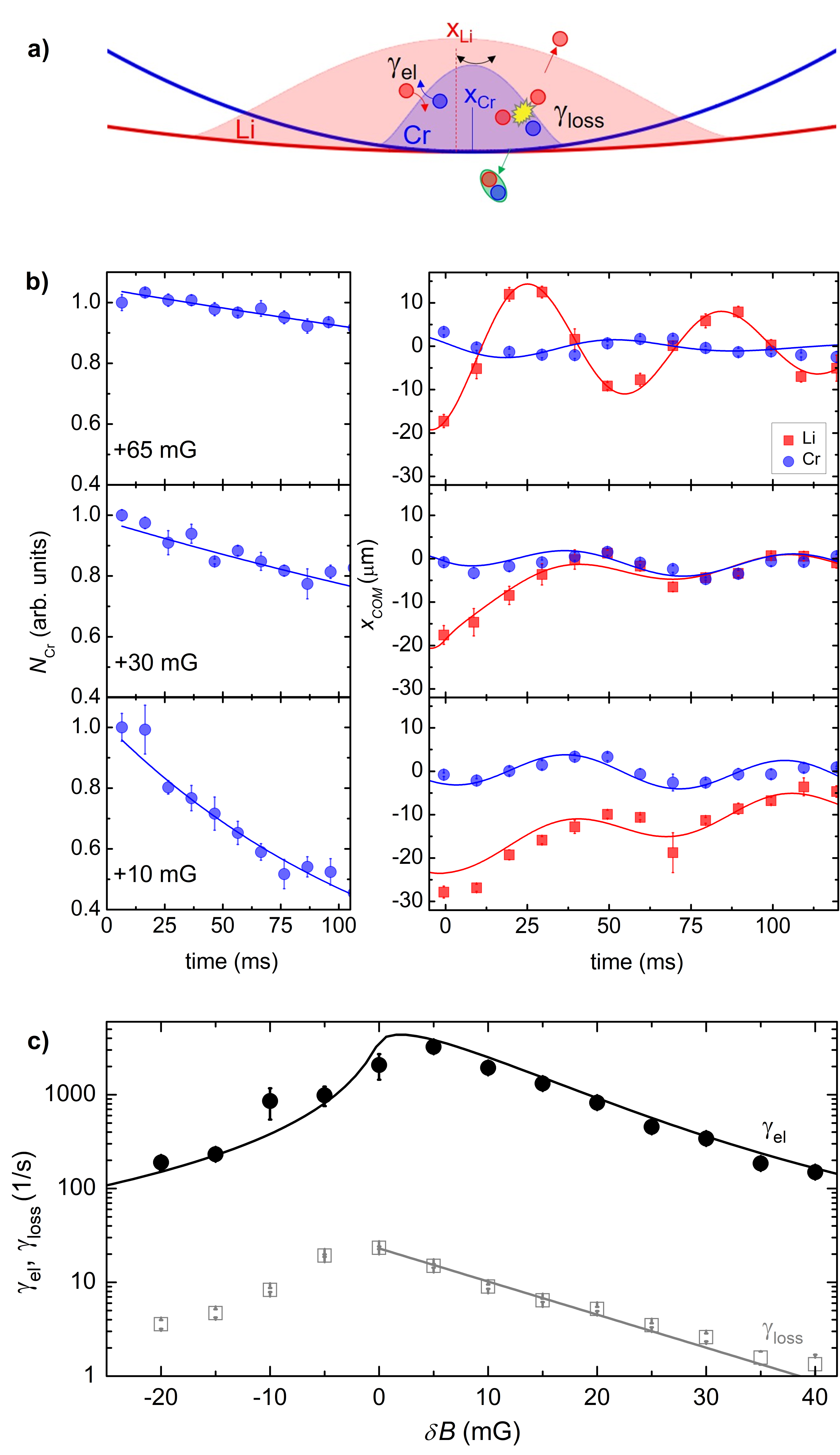}
    \caption{Elastic and inelastic scattering properties of resonantly-interacting Li-Cr Fermi mixtures. (a) Sketch of the experimental configuration employed to simultaneously extract $\gamma_\mathrm{loss}$ from the minority Cr number evolution, and $\gamma_\mathrm{el}$ from  the Li and Cr COM dynamics through the analysis developed in Refs.~\cite{PhysRevA.60.4734,PhysRevLett.87.173201,PhysRevLett.89.053202,FFerlaino_2003}, see text for details. (b) Examples of the observed dynamics, recorded at different magnetic-field detunings specified in the legend, of the chromium atom number (left panels), and of the COMs of both Cr (blue cyrcles) and Li (red squares) components (right panels). Lines are best-fits to the data of the models described in the text, through which we extract $\gamma_\mathrm{loss}$ and $\gamma_\mathrm{el}$, respectively. 
    (c) Experimentally-determined rates for elastic (black cyrcles) and inelastic (empty squares) collision events per Cr atom, as a function of the magnetic field detuning $\delta B$. Black and gray curves are best fits of $\gamma_\mathrm{el}$ and $\gamma_\mathrm{loss}$, respectively, to the theoretically-expected trends, see text for details.}
    \label{fig_hydro}
\vskip -20pt
\end{figure}

The starting point of all our experiments is a weakly-interacting, spin-polarized Fermi mixture of $^6$Li and $^{53}$Cr atoms produced through an all-optical protocol, detailed in Ref. \cite{PhysRevA.106.053318}, within a bichromatic optical dipole trap (BODT).
The latter is realized by two overlapped infrared (IR) and green laser beams, propagating in the horizontal plane along the $x$ direction, and it allows us to adjust the radial confinement of the two components almost independently, owing to the different Li and Cr polarizabilities to the two BODT lights \cite{PhysRevA.106.053318}. 
The gravitational sag between the two clouds is compensated upon application of a weak, vertically-oriented magnetic-field gradient $\partial_z B$ of about 1.5 G$/$cm.
The axial confinement is provided by the curvature of our magnetic field coils, employed to generate bias fields of up to 1.5 kG, and yielding a trapping potential along $x$ six times tighter for chromium than for lithium, see sketch in Fig.~\ref{fig_hydro}(a).
Although atom number, temperature and degree of degeneracy of the two species can be widely tuned by adjusting the (absolute and relative) power of the two BODT beams \cite{PhysRevA.106.053318}, our experiments typically start  with a Fermi gas of about $1.5 \times 10^5$ Li atoms at $ T/T_F= 0.25$ with $T=130(20)\,$nK coexisting with a moderately degenerate sample of $0.8 \times 10^5$  Cr atoms at $ T/T_F \simeq 0.5$ and $T=220\,$nK, with peak densities of about $10^{12}\,$cm$^{-3}$ for both components (here $T_F$ denotes the Fermi temperature of a harmonically-trapped Fermi gas).

Resonant tuning of Li-Cr interaction, as well as magneto-association of atom pairs into LiCr dimers via magnetic-field sweeps, are enabled by the presence of several interspecies FRs that we recently identified, see Ref. \cite{PhysRevLett.129.093402}.
In particular, we focus on two specific $s$-wave FRs that occur in the Li$|1\rangle$-Cr$|1\rangle$ and Li$|2\rangle$-Cr$|1\rangle$ scattering channels (Zeeman levels are labeled in order of increasing energy), located at high fields around $1414\,$G and $1461\,$G, respectively, see Fig.~\ref{fgr:OverviewMerged}(b).
Our choice is motivated by the fact that these two features exhibit the largest magnetic-field width $\Delta B\sim 0.48\,$G available in our mixture, combined with zero or negligible two-body loss rates. Most importantly for the present study, these FRs are induced by hyperfine coupling with the least-bound, rotationless vibrational level of the $X^6\Sigma^+$ ground-state Born-Oppenheimer potential, see Fig.~\ref{fgr:OverviewMerged}: As such, association of such \textit{sextet} dimers represents an excellent starting point to subsequently implement efficient STIRAP transfer to the absolute ground state.
However, in light of the high-field location of the selected FRs and of their narrow nature, as a crucial, preliminary step towards  ultracold LiCr dimers, we first characterize the  parent atomic mixture in the resonantly-interacting, quantum-degenerate regime, going beyond our previous work \cite{PhysRevLett.129.093402}. In particular, exploiting our improved control over magnetic fields in the setup (details in Appendix \ref{Appendix:FieldStabilization}), we accurately locate the FR pole, across which we extract both the elastic and inelastic collision rates per minority Cr atom, denoted $\gamma_{\mathrm{el}}$ and $\gamma_\mathrm{loss}$, respectively.

We simultaneously measure these two observables as a function of the magnetic field detuning $\delta B\!=\!B-B_0$, where $B_0$ is the FR pole, through the study of in-trap collective-mode dynamics, see sketch in Fig.~\ref{fig_hydro}(a). Our probing method is based on the simultaneous monitoring, as a function of time, of the atom numbers $N_{\mathrm{Cr}}$ and $N_{\mathrm{Li}}$, as well as of the axial sloshing of both species in terms of their center-of-mass (COM) coordinates. Fitting exponential decays to the minority Cr atom number yields $\gamma_{\mathrm{loss}}$ associated with the dominant three-body Li-Li-Cr recombination process, see Appendix \ref{Appendix:K3coeff&FRpoles}. The elastic collision rate $\gamma_{\mathrm{el}}$ is instead extracted from the analysis of the COM evolution of the two mixture components via a two-coupled harmonic oscillator model \cite{PhysRevA.60.4734,PhysRevLett.87.173201,PhysRevLett.89.053202,FFerlaino_2003}, see Fig.~\ref{fig_hydro}(b) and Appendix \ref{Appendix:CollRates} for details of the model. Fitting the COM position of the atomic clouds $x_{\textrm{COM}}$  as a function of time and taking into account the observed atom number loss, $\gamma_{el}$ can be extracted reliably.

The measurement starts by preparing a non-interacting Li-Cr mixture within the sole IR trapping beam near the zero-crossing of the selected FR, at an initial detuning $\delta B\!\sim$+0.48\,G. After selectively displacing the Li cloud slightly out of the axial potential minimum at $x\!=\!0$, while leaving Cr unaffected, we quickly set $\delta B$ to variable values around the FR pole, see Fig.~\ref{fig_hydro}(a). We then monitor the subsequent number and COMs evolution as a function of time.  
At the initial detuning, i.e. under non-interacting conditions, the atom losses are negligible, lithium  undergoes small, undamped oscillations with single oscillation frequency of about 17\,Hz. Simultaneously, chromium, which features a bare trapping frequency of 14\,Hz, is found at rest at all evolution times. As the resonance is approached,  effects of both inelastic and elastic processes become more pronounced. Examples of the observed dynamics at various $\delta B\!>\!0$ are presented in Fig.~\ref{fig_hydro}(b) in order of increasing interaction strength from top to bottom. Here we show the minority chromium number $N_{\mathrm{Cr}}$ (left panels) and COM coordinates for both species (right panels), as a function of time. The Cr number evolution, always well fitted by a single exponential decay (see solid lines in the left panels of Fig. ~\ref{fig_hydro}(b)), reveals a progressively reduced lifetime (increased loss rate), as the resonance pole is approached from above, $\delta B\!\rightarrow\!0^+$.
Parallel to this, the COMs of the two mixture components exhibit more interesting dynamics, that qualitatively change close enough to the FR pole. In particular, for $\delta B\gtrsim$  30\,mG 
(small interaction strength) we observe a damping of the Li COM oscillations -- accompanied by weak damped oscillations of the Cr cloud at its bare axial frequency. This damping becomes progressively more pronounced as $\delta B$ is reduced, consistently with the expected behaviour in the collisionless  regime~\cite{PhysRevA.60.4734,PhysRevLett.87.173201,PhysRevLett.89.053202,FFerlaino_2003}.
For $\delta B\lesssim$ 30 mG, instead, the barycenters of the two clouds exhibit weakly-damped, in-phase oscillations, characterized by one single frequency, intermediate between the unperturbed Li and Cr ones. This observation is consistent with a collisionally-hydrodynamic behavior of our Fermi mixture near the FR center, expected to arise in the strongly-interacting region, once $\gamma_\mathrm{el}$ greatly exceeds the axial trap frequencies~\cite{PhysRevA.60.4734,PhysRevLett.87.173201,PhysRevLett.89.053202,FFerlaino_2003}, see Appendix~\ref{Appendix:CollRates} for more details.
Despite the COM evolution drastically changes between the collisionless and the hydrodynamic regimes, it is well reproduced at all detunings by the coupled oscillators model, with $\gamma_{\mathrm{el}}$ as the only free parameter, see best fit curves in right panels of Fig.~\ref{fig_hydro}(b).

The main results of this characterization are summarized in Fig.~\ref{fig_hydro}(c), that shows the experimentally-determined $\gamma_\mathrm{el}$ (full circles) and  $\gamma_\mathrm{loss}$ (empty squares) as a function of the magnetic field detuning across the resonance region. 
The inelastic loss rate, for $\delta B\!\geq\!0$, exhibits an exponential growth (gray line in Fig.~\ref{fig_hydro}(c)) as the resonance pole is approached, in qualitative agreement with the trend expected for the three-body recombination rate coefficient near a narrow FR \cite{PhysRevLett.120.193402}, see Appendix \ref{Appendix:K3coeff&FRpoles} for more details. After reaching a maximum value at $\delta B \!\sim\!0$, the extracted $\gamma_\mathrm{loss}$ is found to progressively drop as the field is further decreased, consistently with previous observations on other atomic systems \cite{PhysRevLett.120.193402}. However, since in this $\delta B\!<\!0$ region Feshbach dimers can be formed and contribute to the detected atom signal, $\gamma_\mathrm{loss}$ cannot be anymore interpreted solely in terms of three-atom recombination processes, and its theoretical analysis goes beyond the scope of the present study.

Contrarily to the inelastic case, $\gamma_\mathrm{el}$ can be analyzed as an elastic Li-Cr scattering rate across the entire region of explored detunings. We obtain the theory estimate as $\gamma_\mathrm{el} =\langle n_{\textrm{Li}}\rangle_{\textrm{Cr}} \langle \sigma(B) v \rangle_T$ \cite{VarennaNotesZaccanti2022}, with $\sigma(B)$ denoting the magnetic-field dependent cross section for Li-Cr collisions, $v$ the relative velocity, $\langle n_{\textrm{Li}}\rangle_{\textrm{Cr}}$ the Li density averaged over the Cr cloud, and $\langle \ldots \rangle_T$ thermal averaging, see Appendix~\ref{Appendix:CollRates} for details. Since the overlap density and temperature of the sample are directly accessible via absorption imaging, and since  $\sigma (B)$ is completely determined by the parameters of our FR \cite{PhysRevLett.129.093402}, see Appendix~\ref{Appendix:CollRates}, we are left with the FR pole location $B_0$ as the only free parameter. The theory fits the experimental $\gamma_\mathrm{el}$ remarkably well across the entire FR region, see solid black line in Fig.~\ref{fig_hydro}(c). This allows us to pinpoint the resonance pole with $\pm$\unit[3]{mG} accuracy.

Finally and most importantly, Fig.~\ref{fig_hydro}(c) data reveal the collisional stability of the Li-Cr mixture under resonant interactions. This is signalled by the elastic rate greatly exceeding the inelastic one over a comparably wide range of magnetic-field detunings across the Feshbach resonance, with a \textit{good-to-bad} collision ratio $\gamma_\mathrm{el}/\gamma_\mathrm{loss}$ found to reach values up to 200.
We ascribe the observed stability to the fact that, in spite of the narrow nature of the FR used \cite{PhysRevLett.129.093402}, three-body processes involving identical fermions are suppressed by anti-bunching due to Fermi statistics of our mixture components  \cite{Petrov2004,RegalPhDThesis}, relative to the bosonic case. As shown in the following section, this is extremely advantageous for the efficient magneto-association of LiCr dimers and, more generally, it is very promising in light of future many-body studies of strongly-interacting Li-Cr Fermi mixtures.

Despite the data on elastic and inelastic properties of our system here presented were taken on Li$|1\rangle$-Cr$|1\rangle$ mixtures across the most promising $s$-wave FR, a similar characterization has been conducted also on Li$|2\rangle$-Cr$|1\rangle$, yielding essentially identical results. 

\section{Production of high phase-space density L\MakeLowercase{i}C\MakeLowercase{r} samples}
\label{Sec:MoleculeFormation}
We now move to investigate the conversion process into LiCr Feshbach molecules via magneto-association. We first characterize the molecule conversion efficiency as a function of the magnetic $B$-field sweep rate across the FR, and then explore how the number of LiCr dimers depends on the initial atomic samples to maximize the final molecule PSD.
In order to experimentally show LiCr dimer formation, count their number, and characterize their density or momentum distribution, here and in the following sections we use negative or positive signals depending on the task at hand, see Appendices~\ref{Appendix:K3coeff&FRpoles} and \ref{Appendix:AbsImgFBmols}. By negative signals we denote atomic loss signals which unambiguously determine the molecule number. This is relatively straightforward in our system, where  two-body losses are absent and three-body recombination is much slower than typical molecule association times, so that only atoms bound into molecules are lost. Positive signals instead
only originate from previously associated atoms, which give rise to a measurable optical density on the atomic Li and Cr imaging
transitions, see Sec.~\ref{Sec:MoleculeProperties_OpenCloseFractions}.  
In our experiment we employ two different schemes to obtain positive signals. The first one relies on Stern-Gerlach separation,
where the molecule cloud is spatially resolved from the atomic ones thanks to a different acceleration and a sufficiently long time of flight (TOF). 
The second method exploits fast (few hundred \textmu s-long) RF transfers of unpaired Li and Cr atoms to dark states which do not interact with the imaging light.
Only positive signals allow access to further information about the molecule sample, such as thermodynamic quantities.

\subsection{Two-body adiabaticity}
\label{Sec:MoleculeFormation_LandauZener}
\begin{figure}[t]
\centering
  \includegraphics[width=\columnwidth]{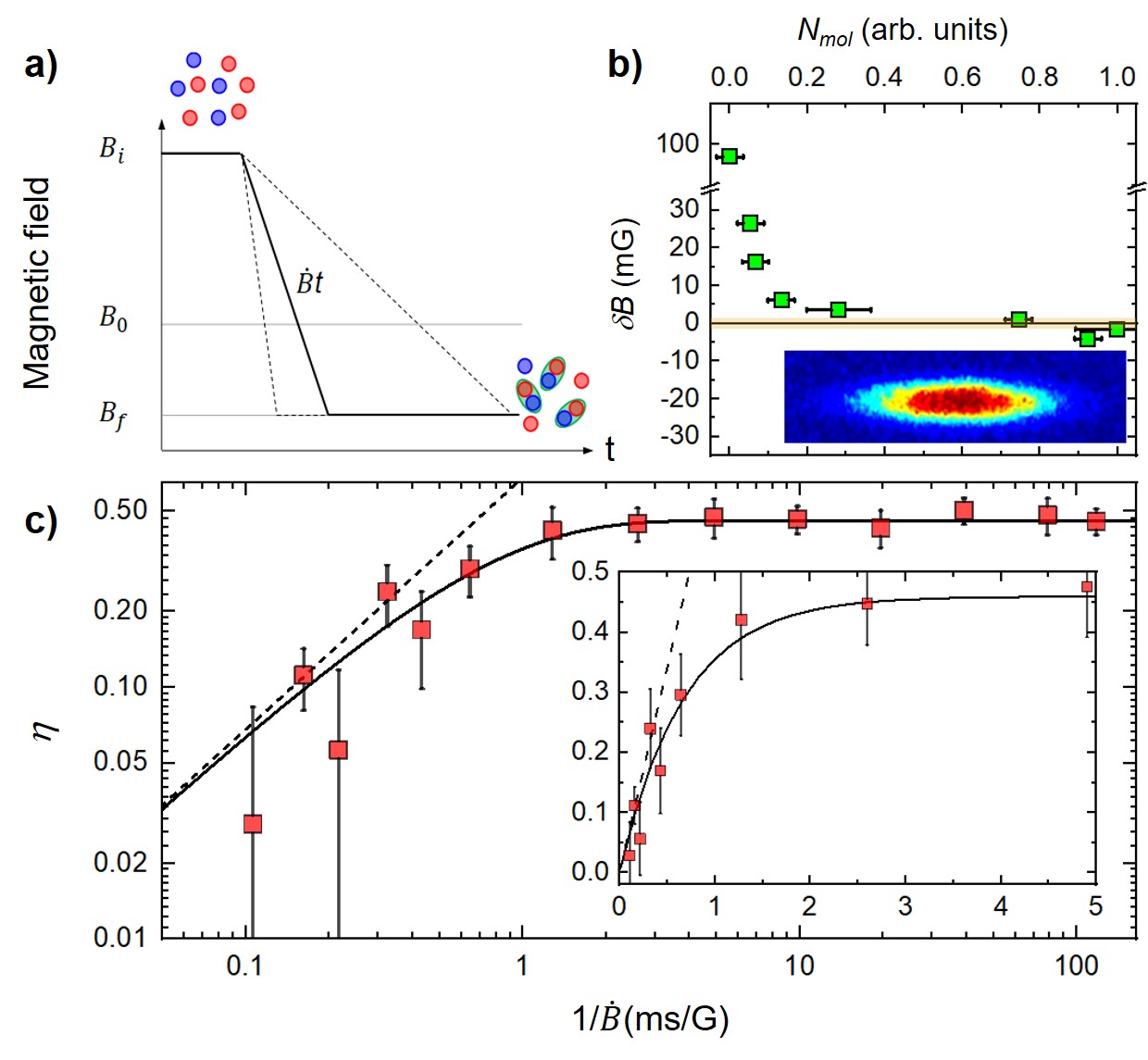}
  \caption{Magneto-association efficiency and two-body adiabaticity. (a) Sketch of the experimental sequence: the magnetic field is swept from $B_i$ across the FR pole $B_0$ down to $B_f<B_0$ with variable rate $\dot{B}$, after which 
 the number of created molecules is measured through either positive or negative signals, see text for details. (b) Normalized number of molecules formed via magneto-association as a function of $B_f$ across the FR for fixed $\dot{B}=90\,\textrm{ms/G}$. The inset is an example of a positive signal of LiCr via direct absorption imaging with Li light after a 2\,ms TOF. (c) Association efficiency $\eta$ referenced to the Cr minority component as a function of the inverse ramp speed $1/\dot{B}$. The solid black line shows the best functional fit of Eq.~(\ref{eq:LZmodel}) to the data and the dashed line its approximation in the fast sweep limit.}
  \label{fgr:LZwithInset}
\end{figure}

Magneto-association is the adiabatic conversion of scattering atom pairs into weakly-bound molecules, induced by a magnetic-field sweep across the FR pole from the attractive ($\delta B\!>\!0$) to the repulsive side ($\delta B\!<\!0$), see Fig.~\ref{fgr:LZwithInset}(a). The efficiency of this process depends on the Feshbach resonance parameters, the conditions of the parent atomic gases, and the magnetic-field sweep rate \cite{RevModPhys.78.1311}. We exploit a positive molecule signal obtained from direct absorption imaging to reveal the sensitivity of the associated molecules to the final magnetic field detuning, see Fig.~\ref{fgr:LZwithInset}(b). We typically observe a sharp rise from zero background to saturation in a $B$-field  region of a few tens of mG. Alternatively, leveraging on the stability of our Fermi mixtures, a negative signal can be used to quantify the number of associated molecules. Indeed, after a short hold time following the $B$-field sweep, only atoms which were associated into dimers are (entirely) lost, and the conversion efficiency is nothing but the fractional loss between the initial and final atom numbers, see Appendix \ref{Appendix:K3coeff&FRpoles}.  
We measure the association efficiency as the fractional loss of the Cr minority component as function of the inverse magnetic-sweep rate $1/ \dot{B}$, see Fig.~\ref{fgr:LZwithInset}(c). We start the experiment by preparing the Li$|1\rangle$-Cr$|1\rangle$ mixture about +100\,mG above the resonance. Li$|1\rangle$ is degenerate with $T/T_F=0.20(5)$ at $T=170\,$nK and a peak density of $1.3(2)\times10^{12}\,$ cm$^{-3}$, while Cr$|1\rangle$ is essentially thermal $T/T_F\simeq 1$ at $T=240\,$nK and a peak density of $0.74(3)\times10^{12}\,$cm$^{-3}$. The magnetic field is swept to $\delta B\!<0$ with variable ramp speed, after which we measure the number of associated molecules. 

As for other experiments, our results are well captured by a Landau-Zener model \cite{PhysRevLett.91.080406,Regal2003,PhysRevLett.94.120402,PhysRevLett.97.180404,PhysRevLett.102.020405,PhysRevA.87.012703}. According to Ref.~\cite{PhysRevA.87.012703}, we fit the following functional form to the data:
\begin{equation}
\label{eq:LZmodel}
\eta=\eta_0 \left( 1-\exp\left(-\frac{\Gamma \langle n_{\textrm{Li}} \rangle_{\textrm{Cr}}}{\eta_0}\frac{1}{ \dot{B}}\right)\right).
\end{equation}
Here $\eta_0$ represents the saturated efficiency, $\langle n_{\textrm{Li}} \rangle_{\textrm{Cr}}$ is the density of the majority component averaged over the density of the minority one, and $\Gamma$ is only function of the collision parameters
\begin{equation}
\label{LZGamma}
\Gamma=\frac{(2 \pi)^2 \hbar}{m} a_{bg} \Delta B,
\end{equation}
where $a_{bg}=41.48(3)\,\textrm{a}_0$ is the background s-wave scattering length, $\Delta B=0.48\,$G is the magnetic field width, and $m$ is the reduced mass. While the adiabatic regime ($\eta \simeq \eta_0$) depends on the overall PSD overlap of the atomic mixture and lacks an analytical description to date, the fast sweep regime, where $\eta\simeq \Gamma \langle n_{\textrm{Li}} \rangle_{\textrm{Cr}} / \dot{B}$, can be straightforwardly tested against the experimental results. From a functional fit of Eq.~(\ref{eq:LZmodel}) with an observed Li$|1\rangle$ density averaged over the Cr$|1\rangle$ cloud of $\langle n_{\textrm{Li}} \rangle_{\textrm{Cr}}=0.62(5)\times10^{12}$ cm$^{-3}$ we derive: $\Gamma_{\textrm{exp}} = 1.1(1)\times 10^{-12}$  cm$^3$ G/ms, see solid line in Fig.~\ref{fgr:LZwithInset}(c). The dashed line marks the linear, fast-sweep regime. The fitted $\Gamma_{\textrm{exp}}$ is in satisfactory agreement with the theory prediction of $\Gamma_{\textrm{th}} = 0.479 \times 10^{-12}$ cm$^3$ G/ms. The mismatch of about a factor of 2 might be due to an underestimation of the overlap density during molecule association, where the mixture experiences strong attractive interactions. We stress that, as expected from the extremely favourable mixture stability shown in Sec.~\ref{Sec:FermiMixAndProcedures}, we can perform magneto-association with field rates slower than the two-body adiabatic criterion by more than two orders of magnitude without affecting efficiency and molecule number.

\subsection{Optimization and molecule PSD}
\label{Sec:MoleculeFormation_Optimization}

\begin{figure}[t]
\centering
  \includegraphics[width=\columnwidth]{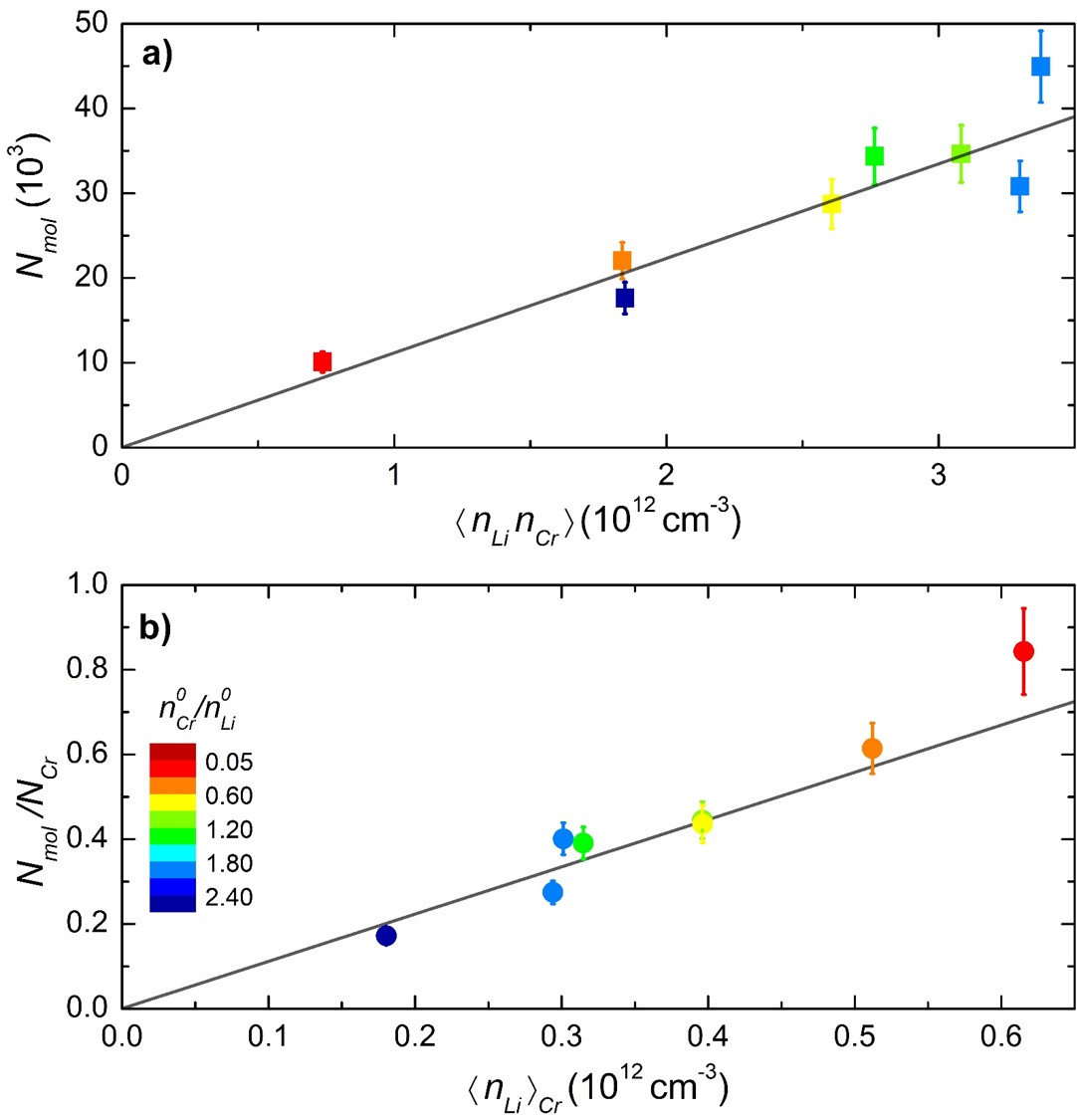}
  \caption{(a)  Number of associated molecules $N_{\textrm{mol}}$ as function of the Li-Cr pair density $\langle n_{\textrm{Li}} n_{\textrm{Cr}}\rangle$. 
  The black line shows a linear fit to the data with zero intercept. (b) Magneto-association efficiency with respect to Cr as a function of the Li density averaged over the Cr cloud $\langle n_{\textrm{Li}} \rangle_{\textrm{Cr}}$. 
  The black line shows a linear fit to the data with zero intercept. In both panels the same color code indicates the corresponding imbalance between the peak densities of Cr and Li atomic clouds.}
  \label{fgr:AdiabaticConversion}
\end{figure}

In order to identify the best working conditions, we characterize the molecule association efficiency as a function of the Li-Cr density overlap, deep in the two-body adiabatic regime ($1/\dot{B}=90\, $ms/G). This is accomplished by only varying the number of atoms initially loaded into the BODT with fixed evaporation and final trap parameters. In this experiment we spatially resolve atomic and molecular clouds thanks to Stern-Gerlach separation and simultaneously count the number of Li, Cr and LiCr. Fig.~\ref{fgr:AdiabaticConversion}(a) shows the associated molecule number as a function of the pair density $\langle n_{\textrm{Li}} n_{\textrm{Cr}}\rangle=\int n_\textrm{Li} n_\textrm{Cr}d\bold{r}$. On general grounds, neglecting atom-molecule thermalization, the molecule number depends on the PSD overlap between the parent atomic clouds during association \cite{PhysRevA.87.012703,PhysRevA.87.053604}. However, for Maxwell-Boltzmann gases at constant temperature, this is proportional to $\langle n_{\textrm{Li}} n_{\textrm{Cr}}\rangle$. Since for this experiment $T_{\textrm{Cr},\textrm{Li}}$ are kept constant, with $T_{\textrm{Cr}}/T_{F,\textrm{Cr}} \geq 0.5$ and  almost constant 
$T_{\textrm{Li}}/T_{F,\textrm{Li}} = 0.30(5)$,
we expect this relation to hold to first approximation. Indeed, the experimental data are well captured by a linear fit as shown by the solid black line in Fig.~\ref{fgr:AdiabaticConversion}(a).
    
From the same experimental data we also extract the conversion efficiency $\eta$ as a function of $\langle n_\textrm{Li} \rangle_\textrm{Cr}$, see Fig.~\ref{fgr:AdiabaticConversion}(b). We stress that, although Cr atoms are always lower in number compared to Li atoms, except for the leftmost point of Fig.~\ref{fgr:AdiabaticConversion}(b) with lowest Li density, their peak density can overcome the Li one. Hence, through this characterization, we explore different regimes in which the majority component is locally either Cr or Li, see color map. Also in this case, $\eta$ exhibits a linear increase with $\langle n_\textrm{Li} \rangle_\textrm{Cr}$, showing no saturation effects up to the highest densities explored herein. Comparing Figs. \ref{fgr:AdiabaticConversion}(a) and \ref{fgr:AdiabaticConversion}(b), we see that intermediate mixture imbalance enables the production of the largest molecule samples, of nearly $5\times10^{4}$ molecules, and a comparatively good conversion efficiency of about $40\%\div50\%$. These results are in line with the value reported on $^6$Li$^{40}$K Feshbach molecules \cite{PhysRevLett.102.020405,PhysRevA.81.043637} and higher than the $30\%$ efficiency shown on $^{40}$K$^{161}$Dy \cite{PhysRevResearch.5.033117}. However, our data also show a maximum conversion efficiency of about $80\%$, at the expense of a lower but still considerable molecule number of $2\div3\times10^4$. This regime is comparable with results on deeply-degenerate homonuclear spin-mixtures of $^{40}$K \cite{PhysRevLett.94.120402}, where both components shared the same degree of degeneracy. The ability to create large molecule samples with tunable density imbalance of the parent mixture may help in the future to sympathetically cool molecules down to degeneracy with leftover Cr atoms, which do not limit the molecule lifetime at our typical densities, see Sec. \ref{Sec:MoleculeLifetime}.

As shown in Fig.~\ref{fgr:AdiabaticConversion} the properties of the molecular gas sensitively depend upon the initial atomic mixture. After careful optimization, we study the ballistic expansion of Feshbach dimers after variable time of flight via the Stern-Gerlach separation method, and compare it with the expansion of the atomic clouds. After the switch-off of the BODT, atoms and molecules expand into the magnetic saddle potential generated by our coils. 
We obtain our record PSD starting magneto-association with atom numbers $N_\textrm{Li}=3 \times 10^5$ and $N_\textrm{Cr}=1 \times 10^5$ and temperatures $T_\textrm{Li}=0.15 T_F=70\,\textrm{nK}$ and $T_\textrm{Cr}=0.5 T_F=180\,\textrm{nK}$. From this starting condition we obtain $36(4) \times 10^3$ molecules at 180\,nK, with peak spatial density of $0.75(10)\times 10^{12}\,\textrm{cm}^{-3}$, and peak PSD of $n_{\textrm{LiCr}} \lambda_{\text{dB}}^3=0.12(2)$, where $\lambda_{\text{dB}}=\sqrt{2 \pi \hbar^2/(m k_\textrm{B} T)}$ is the corresponding de Broglie wavelength. Our result compares well with the highest reported PSDs in other mass-imbalanced Fermi mixture experiments \cite{PhysRevA.94.062706,PhysRevResearch.5.033117}, and even higher LiCr PSDs may be obtained by adding a crossed BODT \cite{PhysRevA.106.053318}, not employed in the present work.
Although the magneto-association ramp is deep in the adiabatic regime for the two-body problem, the molecular sample has not reached full thermal equilibrium. This last point, together with the possibility to further evaporate Cr thereby sympathetically cooling LiCr will be the subject of a future study.

\section{Properties of Feshbach dimers}
\label{Sec:MoleculeProperties}
In this section we directly measure the magnetic dipole moment $\mu_b$ of the newly created LiCr Feshbach molecules and reveal its dependence upon the magnetic field detuning $\delta B$ from the FR pole, following a strategy similar to Ref. \cite{PhysRevResearch.5.033117}. Furthermore, we devise a new all-optical method that allows for an independent measurement of the so-called open-channel fraction of the Feshbach state and of its binding energy.

\subsection{Magnetic dipole moment}
\label{Sec:MoleculeProperties_MagneticMoment}
\begin{figure}[t]
\centering
  \includegraphics[width=\columnwidth]
  {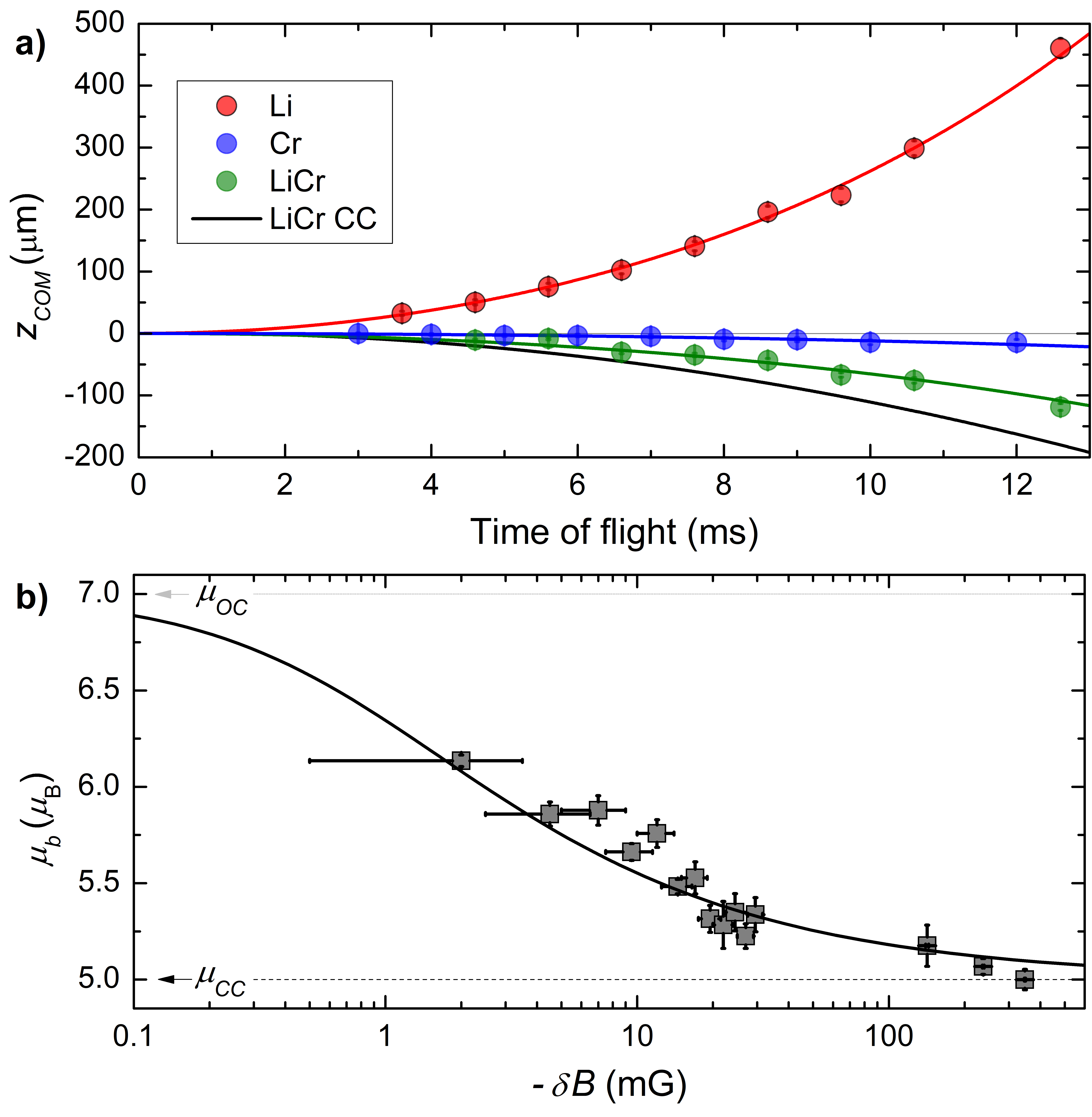}
  \caption{(a) Vertical center-of-mass trajectories for atoms and molecules in time of flight. The red, blue, and green circles show the evolution of the COM in the vertical direction as a function of TOF for Li, Cr, and LiCr, respectively. The corresponding lines with matching color code show the fitted theory curves. The black line represents the expected trajecory  for a purely  closed-channel molecule ($Z=1$). (b) Magnetic dipole moment $\mu_b$ of LiCr Feshbach molecules as a function of the absolute $B$-field detuning from the FR pole. The black curve represents the theory prediction based on Eqs.~(\ref{eq:HFtheorem}) and (\ref{eq:Closed-channel Fraction}) with only the resonance position $B_0$ as fitting parameter. The dotted and dashed lines mark, respectively, the magnetic moments of the scattering atom pair $\mu_{OC}$ ($Z=0$), and of the closed-channel molecule $\mu_{CC}$ ($Z=1$).}
  \label{fgr:MagMoment}
\end{figure}

We first accurately measure $\mu_b$ at fixed $\delta B\!=\!-4\,$mG to reveal a deviation with respect to the the atomic values of Li and Cr, which at this bias field are respectively $\mu_{\textrm{Li}}=1\,\mu_\textrm{B}$ and $\mu_{\textrm{Cr}}=6\,\mu_\textrm{B}$ within a few per mille. To this end we perform a Stern-Gerlach type experiment and record the temporal evolution of the vertical center-of-mass positions $z_{\textrm{COM}}$ of Li, Cr, and LiCr under the application of a vertical magnetic field gradient $\partial_z B$. We prepare the LiCr sample under the optimal conditions found in Sec. \ref{Sec:MoleculeFormation} and switch off the BODT immediately after molecule creation, letting the particles expand under the combination of the gravitational and the species-dependent magnetic forces. For each component, the COM acceleration is $g + \mu_X \, \partial_z B / m_X$, where $g$ is the gravitational acceleration, $\mu_X$ is the magnetic dipole moment of species $X$, and $m_X$ the corresponding mass. The magnetic field gradient $\partial_z B$ is calibrated on the observed $z^{\textrm{Li,Cr}}_{\textrm{CM}}$, and $\mu_b$ is obtained from from a single-parameter fit to the LiCr trajectory, see data and fit in Fig.~\ref{fgr:MagMoment}(a). The extracted value $\mu_b=5.85(5)\, \mu_\textrm{B}$, clearly resolved from both the atomic counterparts, markedly differs from the magnetic moment of the (closed-channel) molecule state inducing the FR $\mu_{CC}=5\,\mu_\textrm{B}$, pointing to a sizable admixture with the open-channel wavefunction, with $\mu_{OC}=7\,\mu_\textrm{B}$.

In order to better understand our observations, it is instructive to express the Feshbach dimer wavefunction as a superposition of open- and closed-channel components \cite{RevModPhys.82.1225,RevModPhys.78.1311,VarennaNotesZaccanti2022}: 
\begin{equation}
\label{FeshbachState}
|\psi_b(R)\rangle=\sqrt{1-Z} \phi_{OC}(R) |OC\rangle+ \sqrt{Z} \phi_{CC}(R) |CC\rangle,
\end{equation}
where the unit-normalized radial wavefunctions $\{\phi_{OC}(R), \phi_{CC}(R) \}$ refer to the open- and closed-channel components with spin states $\{|OC\rangle,|CC\rangle\}$, respectively. The closed-channel fraction $Z$, bound to be $0\!\leq\!Z\!\leq\!1$, determines the character of the Feshbach molecule: from open-channel like for $Z \rightarrow 0$, to closed-channel like for $Z \rightarrow 1$.  The Feshbach dimer has energy $E_b$ and magnetic moment $\mu_b=\partial_B E_b$. These, if referenced to the scattering continuum, define the binding energy $\epsilon_b=E_{OC}-E_b$ and the differential magnetic moment $\delta \mu_b=\mu_{OC}-\mu_b$, which are linked to $Z$ via the Hellmann-Feynman theorem \cite{Politzer2018,RevModPhys.82.1225}:
\begin{equation}
\label{eq:HFtheorem}
\mu_b= Z \mu_{CC}+ (1-Z) \mu_{OC}.
\end{equation}

An analytical solution for $Z(\delta B)$ is available in the effective range expansion \cite{10.1093/acprof:oso/9780199661886.003.0003}
\begin{equation}
\label{eq:Closed-channel Fraction}
Z=1-\frac{1}{\sqrt{1 + 4 R^*/(a-a_{bg})}},
\end{equation}
where $a(B)$ is the scattering length, $a_{bg}$  its background value, and $R^*=\hbar^2/(2 m a_{bg} \Delta B \delta \mu)$  the effective range parameter. 
Eqs.~(\ref{eq:HFtheorem}) and (\ref{eq:Closed-channel Fraction}) thus provide the expected magnetic field dependence of $\mu_b$, ranging for LiCr from $\mu_{\textrm{CC}}=5 \mu_\textrm{B}$ for $Z \rightarrow 1$ to $\mu_{\textrm{OC}}=7 \mu_\textrm{B}$ for $Z \rightarrow 0$. 

We reveal and experimentally characterize the field-dependence of $\mu_b$, thereby confirming the paramagnetic nature of LiCr closed-channel molecules. In this case we let the particles expand for a fixed time-of-flight (TOF) after the BODT is switched off and record the $z_{\textrm{COM}}$ of the LiCr cloud as a function of $\delta B$. In order to minimize temporal and spatial variation of the magnetic field experienced by the molecules, we employ a short TOF of 3.5\,ms with a $200\,$\textmu s-long RF cleaning pulse on Cr  right before the imaging pulse. Knowledge of the initial \emph{in situ} position of the atoms and the $B$-field landscape allows us to directly extract $\mu_b(\delta B)$, which is shown in Fig.~\ref{fgr:MagMoment}(b).

A fit of Eq.~(\ref{eq:HFtheorem}) to the data, with $Z(B)$ given by Eq.~(\ref{eq:Closed-channel Fraction}) and $B_0$ as single fitting parameter, yields the solid black line in Fig.~\ref{fgr:MagMoment}(b). We find a good agreement with the experimental results, and ascribe the residual discrepancy to the $B$-field inhomogeneity experienced by the molecules during the TOF. We stress that these results show our high degree of control over the applied magnetic field and, most importantly, the paramagnetic nature of the LiCr closed-channel molecule. Moreover, the observed trend of $\mu_b$, asymptotically approaching $\mu_{CC}=5\,\mu_\textrm{B}$, confirms the FR assignment of our quantum collisional model for Li-Cr developed in Ref.~\cite{PhysRevLett.129.093402} and shows our ability to controllably populate the least-bound vibrational level of the $X^6\Sigma^+$ ground-state potential. This is extremely convenient for future STIRAP transfer to the LiCr absolute ground state, which only requires a change in the vibrational degrees of freedom and circumvents the need for an optically-excited state with sizable overlap with both electronic ground-state multiplicities as in other experiments~\cite{YangPRL20}, see Sec.~\ref{Sec:TheorySTIRAP}.

\subsection{Optical measurement of open-channel fraction and binding energy}
\label{Sec:MoleculeProperties_OpenCloseFractions}

\begin{figure*}[ht]
\centering
  \includegraphics[width=\textwidth]
  {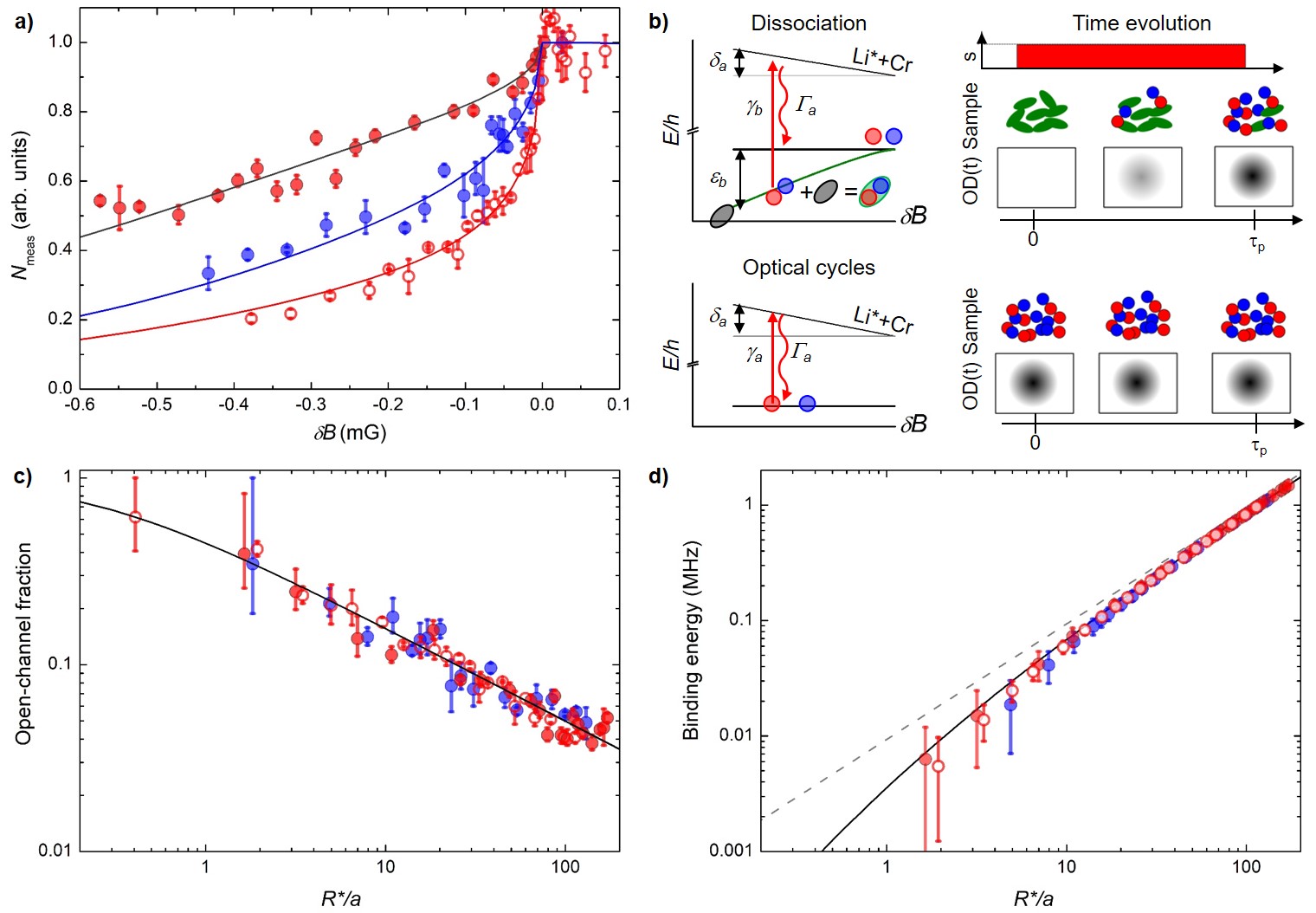}
  \caption{Optically determined open-channel fraction as function of detuning from the resonance. (a) Counted molecule number from in-\emph{situ} absoprtion pictures taken on the Li (red circles) and Cr (blue circles) imaging transitions on pure molecular samples. We show a comparison between a longer (full circles) and shorter (hollow circles) pulse time with the same laser intensity within 10\%. The red full circles are taken with $\tau_P=40\,$\textmu s and $I_0=0.24\,$mW cm$^{-2}$, red open circles with $\tau_P=11\,$\textmu s and $I_0=0.2\,$mW cm$^{-2}$, and the blue circles with $\tau_P=21\,$\textmu s and $I_0=0.76\,$mW cm$^{-2}$. The solid lines show our model prediction Eq.~(\ref{eq:ODSuppressionMolLowOD}) with only a common field offset as fitting parameter and $Z$ given by Eq.\~(\ref{eq:Closed-channel Fraction}). (b) Sketch illustrating the model explained in the main text. The top-left panel shows the dissociation mechanism at rate $\gamma_b$ induced by the imaging laser resonant at $\delta B=0$ with the lithium atomic transition of natural width $\Gamma_a$: a detuning of $\delta_m=\delta_a+\epsilon_b$ is acquired for finite $\delta B<0$. The bottom-left panel shows the optical cycles Li atoms undergo after molecule dissociation at rate $\gamma_a$. The right panels show from top to bottom: the pulse shape of duration $\tau_p$, the time-dependent OD of the initial molecule sample, and, as a comparison, the constant OD of an atomic sample. (c,d) The open-channel fraction and binding energy as a function of $R^*/a$, extracted with our model Eq.~(\ref{eq:ODSuppressionMolLowOD}) from the three datasets in panel (a) (with the same symbol choice), together with the theory predictions given by Eq.~(\ref{eq:Closed-channel Fraction}) and Eq.~(\ref{eq:BindingEnergy}), respectively (solid black lines), see Appendix~\ref{Appendix:AbsImgFBmols} for details.} 
  \label{fgr:OCFractionPanelMathematica}
\end{figure*}

As anticipated in the previous section discussing our magnetic moment measurement (see Fig.~\ref{fgr:MagMoment}), the closed-channel fraction represents an essential feature of the Feshbach dimer. Besides $\mu_b$, $Z$ also determines the collisional stability of Feshbach molecules, as well as the outcome of future STIRAP transfer schemes to deeply-bound states. 
While the former is longest for $Z \rightarrow 0$, the latter is optimal for $Z \rightarrow 1$.
An accurate measurement of $Z(\delta B)$ close to a narrow FR at high $B$-field is technically challenging. Indeed, wiggle spectroscopy \cite{PhysRevLett.95.190404} has limited temporal resolution, RF spectroscopy \cite{Regal2003,PhysRevLett.94.103201} requires high RF power potentially perturbing the $B$-field stabilization, photo-association techniques \cite{PhysRevLett.95.020404} require an optically-excited molecule level and laser light to address it, and, finally, magnetic moment spectroscopy, see previous section and \cite{PhysRevResearch.5.033117}, suffers from field inhomogeneity and temporal constraints.  
Here we overcome these issues by developing a convenient and rather general optical method based on the direct counting of LiCr dimers with atomic imaging light. As detailed in the following, the reduced LiCr imaging signal, observed for increasing $|\delta B|$, allows us to accurately retrieve $Z(\delta B)$ through a fast and inexpensive probing technique. This may find straightforward application to other Feshbach molecule species.

Our measurement starts by preparing a pure sample of LiCr slightly detuned to the $\delta B\!<\!0$ side of the FR. We then quickly ramp the magnetic field to a final detuning $\delta B$, both positive and negative, and take an absorption picture on either Li or Cr atomic lines. The imaging laser frequencies are kept constant and are resonant with the corresponding atomic transitions for $\delta B\!=\!0$. Fig.~\ref{fgr:OCFractionPanelMathematica}(a) reports the observed LiCr number as a function of $\delta B$ both for Li (in red circles) and Cr (blue circles) imaging lights. For Li we further compare a $40\,$\textmu s-long pulse (full circles) with a $11\,$\textmu s-long one (empty circles). Although the apparent number $N_{\textrm{meas}}$ inferred from absorption pictures significantly decreases moving far to the BEC side, the real molecule number is constant in $\delta B$. This point is easily demonstrated in the experiment by dissociating molecules right before taking the absorption picture of the atoms. The magnetic-field dependence of the imaged molecule number has been reported in a number of experiments, and even interpreted as an effective imaging cross section \cite{PhysRevA.87.012703}. However, in the present case the particle number sensitivity on the imaging parameters confutes such an interpretation. 

We explain this observation by developing a simple model, that we describe in the following, and that holds under three main assumptions, see also sketch in Fig.~\ref{fgr:OCFractionPanelMathematica}(b): (i) closed-channel molecules do not directly interact with the atomic imaging light; (ii) Feshbach dimers are dissociated by the imaging light at a rate given by the Fermi's golden rule $\gamma_b=\gamma_a |\langle \psi_b|OC \rangle|^2=\gamma_a (1-Z)$, with $\gamma_a$ denoting the atomic scattering rate; (iii) low-intensity atomic imaging is performed on a cycling transition with atomic scattering rate $\gamma_a(s_a,\Gamma_a,\delta_a)$ dependent on the saturation parameter $s_a\ll 1$, the natural width of the transition $\Gamma_a$, and detuning $\delta_a$. Assumptions (i) and (ii) imply that during the imaging pulse the dimer (column) density decreases as $n_{2D,b}(t)=n^0_{2D,b} e^{-\gamma_b t}$, while the free-atom one grows as:
\begin{equation}
\label{Eq:ImagingTimeDependentAtomDensity}
n_{2D,a}(t)=n^0_{2D,b} (1-e^{-\gamma_b t}). 
\end{equation}
Since (iii) implies that the dissociation products decay back to the same open-channel of the FR, and that the addressed atomic species undergoes cycling transitions until the end of the imaging pulse, from Eq.~(\ref{Eq:ImagingTimeDependentAtomDensity}) we derive the instantaneous optical density (OD)
\begin{equation}
\label{Eq:TimeDependentOD}
OD(t)= \sigma_a n^0_{2D,b} (1-e^{-\gamma_b t}), 
\end{equation}
where $\sigma_a$ is the detuning-dependent, atomic absorption cross section. This is in contrast with the standard case of an atomic sample at the same (column) density, that would feature a constant optical density $OD_a=\sigma_a n^0_{2D,b}$ (see Fig.~\ref{fgr:OCFractionPanelMathematica}(b) sketch). For $\delta B=0$ the molecules are dissociated, the imaging light is resonant with cross section $\sigma_a=\sigma^0_a$, and the optical density is constant and equal to $OD_a=OD^0_a$. 

By integrating Eq.~(\ref{Eq:TimeDependentOD}) over the duration $\tau_p$ of the imaging pulse, it is straightforward to obtain an analytic expression for the ratio between the measured $\overline{OD}(\tau_p)$ at arbitrary detuning $\delta B\!<\!0$ and the atomic $OD^0_a$. While we refer the reader to Appendix \ref{Appendix:AbsImgFBmols} for the general derivation, we here provide a simplified expression holding for $OD^0_a\ll 1$
\begin{equation}
\label{eq:ODSuppressionMolLowOD}
\frac{\overline{OD}(\tau_p)}{OD^0_a}=\frac{1}{1+(2 \delta_a / \Gamma_a)^2}\times \left( 1- \frac{1-e^{-\gamma_b \tau_p}}{\gamma_b \tau_p} \right),
\end{equation}
where $\delta_a(\delta B)=\mu_\textrm{B} \delta B$ is the Zeeman frequency shift of the employed atomic transition. 
The dissociation rate implicitly depends on the detuning of the laser as $\gamma_b=\gamma_a(s_a,\Gamma_a,\delta_b) (1-Z)$, with $\delta_b(\delta B)=\delta_a(\delta B)+\epsilon_b(\delta B)/h$. Being the right-hand side of Eq.~(\ref{eq:ODSuppressionMolLowOD}) independent of $OD^0_a$, it coincides with the suppression of the apparent number of Feshbach molecules, $N_{\textrm{meas}}(\tau_p)/N^0_a=\overline{OD}(\tau_p)/OD^0_a$.

We benchmark our model on the experimental data shown in Fig.~\ref{fgr:OCFractionPanelMathematica}(a) assuming the field dependence of $Z(\delta B)$ from Eq.~(\ref{eq:Closed-channel Fraction}) and the corresponding binding energy $\epsilon_b(\delta B)$ from integration of Eq.~(\ref{eq:HFtheorem}) \cite{10.1093/acprof:oso/9780199661886.003.0003}, $
\epsilon_b(\delta B)=\hbar^2 k_*^2 /2m $ with
\begin{equation}
\label{eq:BindingEnergy}
k_*=\frac{\sqrt{1+4R^*/(a-a_{bg})}-1}{2R^*},
\end{equation}
where the explicit $\delta B$ dependence, via $a(\delta B)$, is omitted for simplicity of notation. Since all parameters appearing in the model are known, we only allow for a precision fit of the resonance location common to all data. The best fit results of our model are shown as solid lines in Fig.~ \ref{fgr:OCFractionPanelMathematica}(a), and reproduce our experimental observation remarkably well, regardless of the imaged species and employed pulse parameters. From this we conclude that our model can be exploited for molecule number calibration with no need for dissociation prior to the imaging pulse, which may introduce systematic heating and excitation \cite{PhysRevLett.92.180402,PhysRevResearch.4.L022019}. In particular, we note that the saturation parameter can be conveniently derived from the reference imaging pulse after the calibration of the camera.

Most importantly, our model, Eq.~(\ref{eq:ODSuppressionMolLowOD}), allows us to retrieve both the open-channel fraction $1-Z(\delta B)$ and the binding energy $\epsilon_b(\delta B)$ of the Feshbach dimers (at all $\delta B$) from their absorption images, without any \emph{a-priori} knowledge of their functional forms. Referring the reader to Appendix \ref{Appendix:AbsImgFBmols} for details, in Figs.~\ref{fgr:OCFractionPanelMathematica}(c) and (d) we show the results of our analysis, plotting the experimentally-determined open-channel fraction and binding energy as a function of $R^*/a$. Notably, extraction of $1-Z$ and $\epsilon_b$ from the three datasets of Fig.~\ref{fgr:OCFractionPanelMathematica}(a) yields consistent results over a wide range of detunings, from the resonant regime to the background limit, despite the different pulse parameters and atomic species employed. Both observables are found to vary over a few orders of magnitude in excellent agreement with the theory predictions based on Eqs.~(\ref{eq:Closed-channel Fraction}) and (\ref{eq:BindingEnergy}), respectively (solid black lines). In particular, we remark how the non-zero open-channel fraction is reflected in the dimer binding energy that deviates, near the resonance pole for $R^*/a\rightarrow0$, from the one of the bare closed-channel molecule $2 \mu_b \delta B$, see dashed line in Fig.~\ref{fgr:OCFractionPanelMathematica}(d). We emphasize that this probing method works remarkably well even at very small detunings from the FR pole. Finally, note also how the data in Figs.~\ref{fgr:OCFractionPanelMathematica}(c) and (d) demonstrate our experimental capability to access the resonant regime with high accuracy, despite the narrow FR at our disposal, a key point for future many-body investigation of Li-Cr mixtures.

\section{Stability of atom-molecule mixtures and long-lived L\MakeLowercase{i}C\MakeLowercase{r} samples}
\label{Sec:MoleculeLifetime}
We now move to address another key question about the experimentally-realized LiCr molecules, namely their stability against inelastic loss processes. 
Since magneto-association results in an optically-trapped mixture of LiCr dimers, together with   unpaired Li and Cr atoms, various inelastic mechanisms may contribute to limit the molecule lifetime. Quite generally, these can arise both from off-resonant scattering of LiCr towards electronically-excited levels -- accidentally induced by the trapping lights employed in the experiment -- as well as from inelastic scattering in dimer-dimer and atom-dimer collisions. 
As the first one-body process is concerned, we find that LiCr molecules are indeed affected by both lights of our BODT.  Specifically, at constant densities, we observe a linear increase of the dimer loss rate with the intensity of both BODT beams, characterized by slopes of $\Gamma_{CC}=397(14)\,\textrm{Hz}/(\textrm{kW}\,\textrm{cm}^{-2})$ and $\Gamma_{CC}=5.9(2)\,\textrm{Hz}/(\textrm{kW}\,\textrm{cm}^{-2})$ for the green and IR light, respectively, see Appendix \ref{Appendix:TrapLightLosses} for details. 
Given the much smaller photo-excitation rate observed for the IR beam, relative to the green one, it is thus advantageous to turn off the latter light when LiCr dimers are created. In this way, at our typical BODT intensities, the one-body lifetime increases from about 2\,ms up to a few tens of ms, allowing us to characterize the stability  of the molecular sample against dimer-dimer (D-D) and atom-dimer (A-D) inelastic collisions. This is fundamental both in light of future optical spectroscopy studies, and of possible implementation of evaporative and sympathetic cooling stages of Feshbach dimers towards quantum degeneracy.

As for three-atom recombination discussed in Sec.~\ref{Sec:FermiMixAndProcedures}, Fermi-Fermi mixtures are expected \cite{Petrov2004,PhysRevA.94.062706,VarennaNotesZaccanti2022} to benefit from Pauli suppression also for D-D and A-D inelastic collisions near the FR pole -- owing to the fact that these always involve at least two identical fermions (either unpaired or paired into a shallow dimer). 
Specifically, the loss rate coefficients $\beta$ of such processes can be substantially decreased, with respect to their off-resonant background value $\beta_{bg}$, as soon as dimers acquire a sizable open-channel fraction, i.e. near the resonance pole, see Sec.~\ref{Sec:MoleculeProperties_OpenCloseFractions}. While this qualitative feature is generic to any Fermi-Fermi mixture, the suppression factor $\beta(R^*/a)/\beta_{bg}$ sensitively depends upon the specific mass ratio between the two constituents. For atom-dimer processes in heteronuclear systems, light-light-heavy inelastic processes are predicted to dominate over the heavy-heavy-light ones in the resonant regime \cite{PhysRevA.94.062706,VarennaNotesZaccanti2022}. For our specific mixture, for instance, Ref. \cite{PhysRevA.94.062706} foresees already for $R^*/a\lesssim$10 a suppression for Cr-LiCr collisions more than one order of magnitude larger than the one for Li-LiCr scattering. 
Moreover, one expects also in the off-resonant regime that lithium-dimer collisions are dominant, given that the  background value of the A-D loss rate coefficient is theoretically given by \cite{PhysRevA.94.062706} $\beta_{\textrm{A-D}}^{bg}\!\sim\!2 h  R_{vdW}/m_{\textrm{A-D}}$, with $R_{vdW}$ the van der Waals length and $m_{\textrm{A-D}}$ the atom-dimer reduced mass, and that $m_{\textrm{Cr-D}}/m_{\textrm{Li-D}}\!\sim\!5$. Also in light of the fact that the Li density exceeds the Cr one in our typical conditions, we observe that collisional losses of LiCr dimers are indeed dominated by Li-D scattering. This is confirmed by a measured Li-to-Cr relative loss of $\Delta N_{\textrm{Li}}/\Delta N_{\textrm{Cr}}=1.93(17)$, and by the fact that, when  unpaired Li atoms are removed from the sample, we do not have any detectable signature of Cr-D or D-D inelastic processes.
 
\begin{figure}[t]
\centering
  \includegraphics[width=\columnwidth]{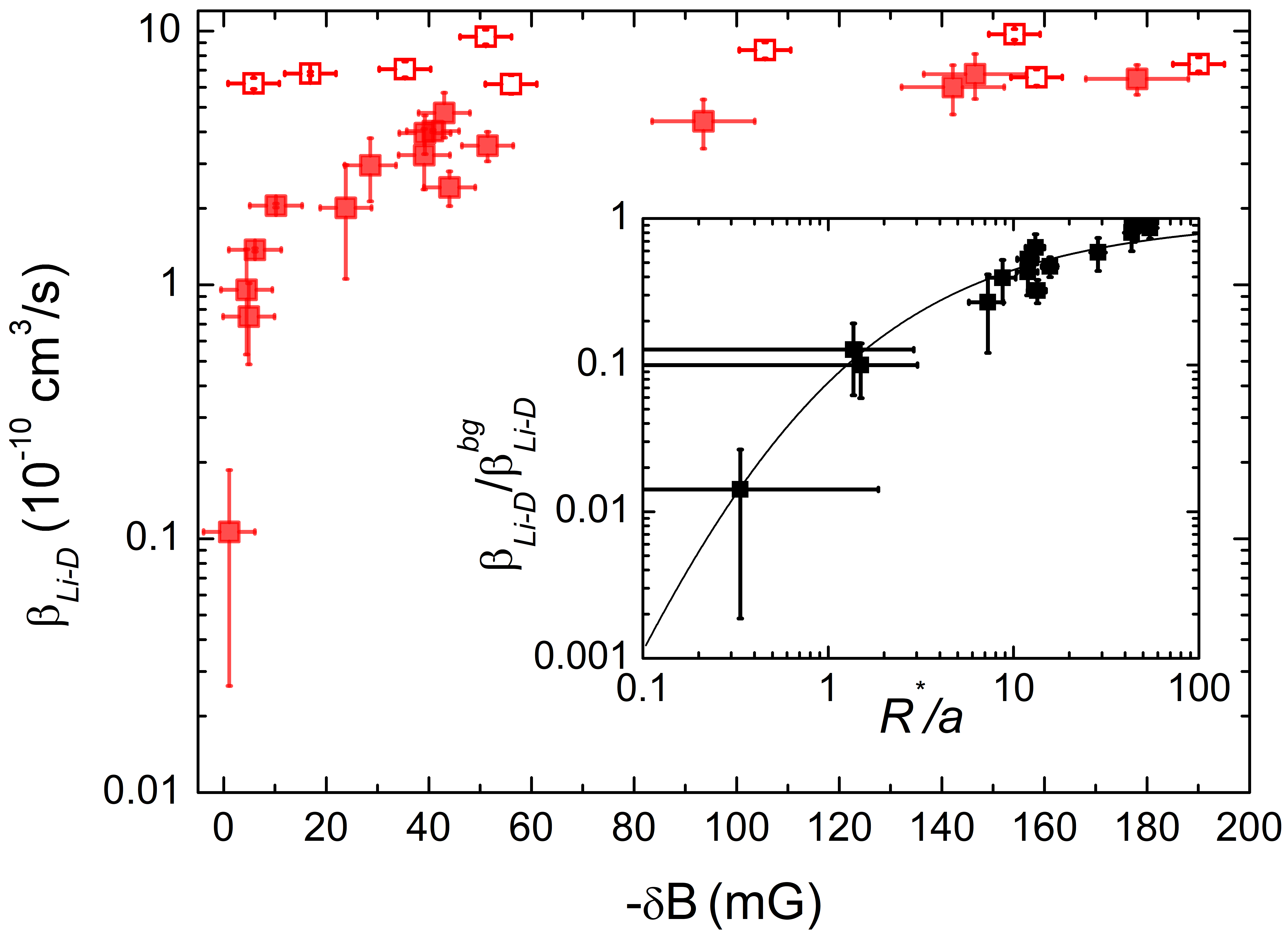}
  \caption{Loss rate coefficient $\beta_{\textrm{Li-D}}$ for inelastic Li-LiCr collisions as a function of magnetic-field detuning. The trend of  $\beta_{\textrm{Li-D}}$, measured for collisions between interacting Li$|1\rangle$ and Li$|1\rangle$Cr$|1\rangle$ molecules (full squares), is compared with the constant, non-resonant one, obtained by studying Li$|2\rangle$-dimer mixtures (empty squares). The inset shows the experimentally-determined $\beta_{\textrm{Li-D}}$, normalized to its background value $\beta_{\textrm{Li-D}}^{bg}\!=7\times 10^{-10}$ cm$^3/$s, as a function of $R^*/a$ (black symbols), together with the theoretically-predicted trend of Ref. \cite{PhysRevA.94.062706}. Vertical error bars of $\beta_{\textrm{Li-D}}$ combine the fit, statistical and systematic uncertainties to the decay data and the estimated mean density $\langle n_{\textrm{Li}}\rangle_\textrm{D}$. Horizontal error bars account for the uncertainty in the magnetic field detuning from the FR pole.
  }
\label{fgr:AtomDimerLosses}
\end{figure}
We characterize the magnetic-field dependence of $\beta_{\textrm{Li-D}}$  varying the endpoint of the association ramp at the Li$|1\rangle$-Cr$|1\rangle$ resonance, and tracing the molecule decay as a function of time, monitored through selective imaging with the Cr light. At each detuning, the fitted exponential decay rate,  corrected for the residual trap-light-induced loss contribution, yields the collisional rate. Dividing the latter by the dimer-weighted Li density $\langle n_{Li} \rangle_{D}$ determined experimentally, we thus extract the inelastic rate coefficient $\beta_{\textrm{Li-D}}$. This is shown in Fig.~\ref{fgr:AtomDimerLosses} as a function of $\delta B$, see full red squares. One can notice how $\beta_{\textrm{Li-D}}$ monotonically decreases down from its off-resonant value as the resonance pole is approached, exhibiting a drop that becomes progressively more pronounced as $\delta B \rightarrow 0^-$. To corroborate the interpretation of our findings in terms of Pauli suppression of collisional losses \cite{PhysRevA.94.062706}, 
we repeat the same measure but transferring the remaining unbound lithium atoms into the non-resonant spin-state Li$|2\rangle$ right after the magneto-association ramp. 
In this  case, where three, rather than two, distinguishable kind of fermionic atoms are involved in the collision, we observe a rate coefficient that is constant at all detunings, see empty squares in Fig.~\ref{fgr:AtomDimerLosses}. This also matches the Li$|1\rangle$-D measured background value, of about $7 \times 10^{-10}\,$cm$^3$/s, which agrees within a factor of two with the theoretical estimate \cite{PhysRevA.94.062706}.
Notably, as shown in the inset of Fig.~\ref{fgr:AtomDimerLosses}, we find excellent agreement between normalized experimental data $\beta_{\textrm{Li-D}}/\beta_{\textrm{Li-D}}^{bg}$ (black squares), and the predicted suppression function for Li-LiCr inelastic collisions given in Ref. \cite{PhysRevA.94.062706} (solid line) as a function of $R^*/a$.
We stress that, in spite of the narrow  nature of our FR, at the smallest detunings here explored we observe up to a factor of $70^{+160}_{-35}$ suppression of collisional losses -- a very promising value in light of future studies of resonantly-interacting atom-dimer mixtures, and of the possible implementation of a final sympathetic cooling stage for LiCr molecules.

\begin{figure}[t]
\centering
  \includegraphics[width=\columnwidth]{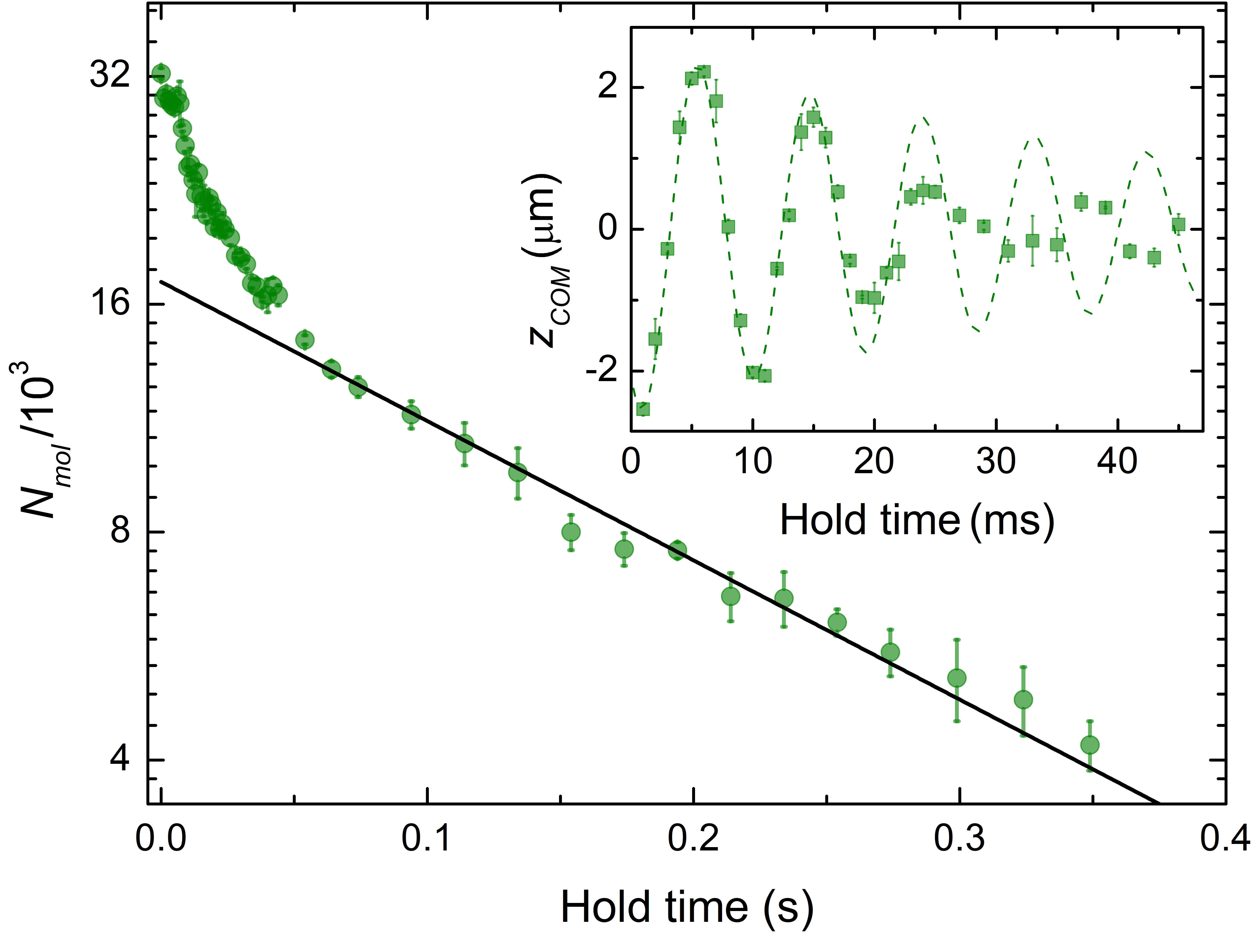}
  \caption{Long-lived pure LiCr samples in the FORT trap at $\delta B\!=-100$ mG. The main panel shows the molecule number (in log-scale) as a function of the hold time, together with an exponential fit of the long-time decay yielding a time constant of 0.24(1) s. For hold times shorter than 50 ms, we observe a faster loss dynamics, which we ascribe to the initial excitation of collective modes of the LiCr sample, see inset, induced by the magneto-association and purification stages.
 }
  \label{fgr:LongLivedMols}
\end{figure}

Finally, we show how the dimer lifetime can be substantially increased at \textit{all} detunings by removing the unpaired atoms and by canceling residual trap-induced off-resonant scattering. To this end,  we make use of an additional 1560\,nm far-off resonant trap (FORT), whose wavelength is found to cause negligible light-induced losses, see Appendix~\ref{Appendix:TrapLightLosses}, consistent with zero within our experimental uncertainty. In this case we first transfer the atomic mixture from the BODT into the FORT following a 100\,ms-long linear ramp, throughout which a constant trap depth for the Cr atoms is maintained. 
We then perform the magneto-association ramp, at the end of which we 
purify the molecule sample from the atomic components. Li atoms are removed through a spin-selective optical blast following a Li$|1\rangle\rightarrow$Li$|2\rangle$ 
RF-transfer. The Cr component is spilled from the FORT by further reducing the trap depth, relying on the fact that the Cr polarizability at 1560\,nm is about a factor 3.5 lower than for LiCr \cite{AtomicPolarizabilities,MolPolarizabilities}. 
While this overall procedure only reduces the initial molecule number by less than a factor of two at about $300\,$nK, it allows for a dramatic gain in lifetime even in the off-resonant regime. This is shown in Fig.~\ref{fgr:LongLivedMols}, where the molecule number is plotted as function of the hold time at a final, large detuning $\delta B\!=-100$ mG. After an initial loss -- ascribable to excitation of collective modes in the LiCr cloud during the magneto-association and atom purification stages, see inset of Fig.~\ref{fgr:LongLivedMols} -- we observe a clean exponential decay of the molecule number, characterized by a time constant of $0.24(1)\,$s. This lifetime, much longer than those measured in our system without purification nor FORT trap (limited to a few tens of ms), exceeds by more than one order of magnitude that reported for LiK molecules \cite{PhysRevLett.102.020405} at similar peak densities, of a few 10$^{11}$ cm$^{-3}$.
The slow exponential decay of LiCr shown in Fig.~\ref{fgr:LongLivedMols} is consistent with residual evaporation dynamics within the relatively low-density molecule cloud, and it appears incompatible with intra-species LiCr  collisional losses, which will be subject of future studies. The present measurement, and analogous ones carried without removal of Cr atoms, yield in all cases upper-bound estimates for the background values of the recombination rate coefficients, $\beta_{\textrm{D-D}}^{bg}$ and $\beta_{\textrm{Cr-D}}^{bg}$, of about $1 \times 10^{-10} \textrm{cm}^3/s$. This, together with the  Pauli suppression of these inelastic processes \cite{PhysRevA.94.062706} -- expected to be even stronger than the one unveiled in Li-D collisions and shown Fig.~\ref{fgr:AtomDimerLosses} -- is very promising for future implementation of collisional cooling of LiCr Feshbach dimers.

\section{Quantum-chemical modeling}
\label{Sec:Theory}

The ultracold samples of $^6$Li$^{53}$Cr Feshbach dimers, thoroughly characterized in the previous sections,  represent an excellent starting point for the   production of a high phase-space density gas of deeply-bound, paramagnetic polar molecules. Yet, LiCr is an almost unexplored species, for which spectroscopic data are completely lacking thus far, except for pioneering electron spin resonance data~\cite{VANZEE1985524}. Its theoretical study, through advanced quantum-chemical simulations, is thus crucial for two reasons: First, to gain information about the ground- and electronically-excited-state properties of the LiCr molecule; Second, to identify suitable pathways to coherently transfer LiCr Feshbach dimers  to the absolute ground state ~\cite{RevModPhys.78.1311}. 
In the following, we employ state-of-the-art \textit{ab initio} quantum-chemical methods to build the complete theoretical model for LiCr. By extending for the first time the portfolio of computational strategies, already applied to alkali-metal or alkaline-earth-metal dimers, to ultracold molecules containing transition-metal atoms, we obtain quantitatively accurate information on both ground and excited electronic states, transition moments, and resulting rovibrational structure and spectra.

\subsection{Electronic and rovibrational structure calculations}

In the first step, we investigate the electronic structure of the LiCr molecule. Potential energy curves and other electronic properties are calculated within the Born-Oppenheimer approximation, including the electron correlation and relativistic effects. The two lowest electronic states, $X^6\Sigma^+$ and $a^8\Sigma^+$, dissociating into ground-state atoms, can be described very accurately using the hierarchy of the coupled cluster methods, including corrections up to the CCSDTQ level~\cite{BartlettRMP07}. Higher-excited electronic states present a challenge, unprecedented in alkali-metal or alkaline-earth-metal molecules, because they involve high electronic spin and large orbital angular momenta of an excited Cr atom. These states are studied using the multireference configuration interaction methods, including single and double excitations and large active spaces. 

We employ large correlation-consistent Gaussian basis sets with basis set cardinal numbers up to 5~\cite{BalabanovPeterson2005,PrascherTCA11} to describe all 27 electrons (3 from Li and 24 from Cr). The scalar relativistic effects are included by the third-order Douglas-Kroll-Hess (DKH) relativistic Hamiltonian~\cite{DouglasKroll,HessPRA86}, while the $R$-dependence of the relativistic spin-orbit coupling is neglected. The atomic basis sets are additionally augmented by a set of bond functions (BF) to accelerate the convergence toward the complete basis set limit (CBS). The counterpoise correction is applied to reduce the basis sets superposition errors within the supermolecule approach to interaction energy calculations~\cite{BoysMolPhys70}. All electronic structure computations are performed with the MOLPRO package of \textit{ab initio} programs~\cite{MOLPRO_brief}. The full triple and quadruple coupled-cluster contributions are computed using the MRCC code embedded in MOLPRO~\cite{MRCC}.  

In the second step, rovibrational eigenstates are calculated numerically with the exact diagonalization of the Hamiltonian for the nuclear motion within the discrete variable representation on the non-equidistant grid~\cite{DVR}. The atomic scattering lengths are computed using the renormalized Numerov propagator~\cite{JohnsonJCP78} with step-size doubling and about 100 step points per de Broglie wavelength.
Theoretical uncertainties are carefully estimated~\cite{GronowskiPRA20} by analyzing the convergence of the electronic structure calculations. More details on the employed computational and theoretical procedures, benchmarks of the used methods and basis sets, and accuracy estimations are presented in the accompanying paper Ref.~\cite{companion}. 

\subsection{Ground-state properties}

\begin{figure}[tb!]
\begin{center}
\includegraphics[width=1\linewidth]{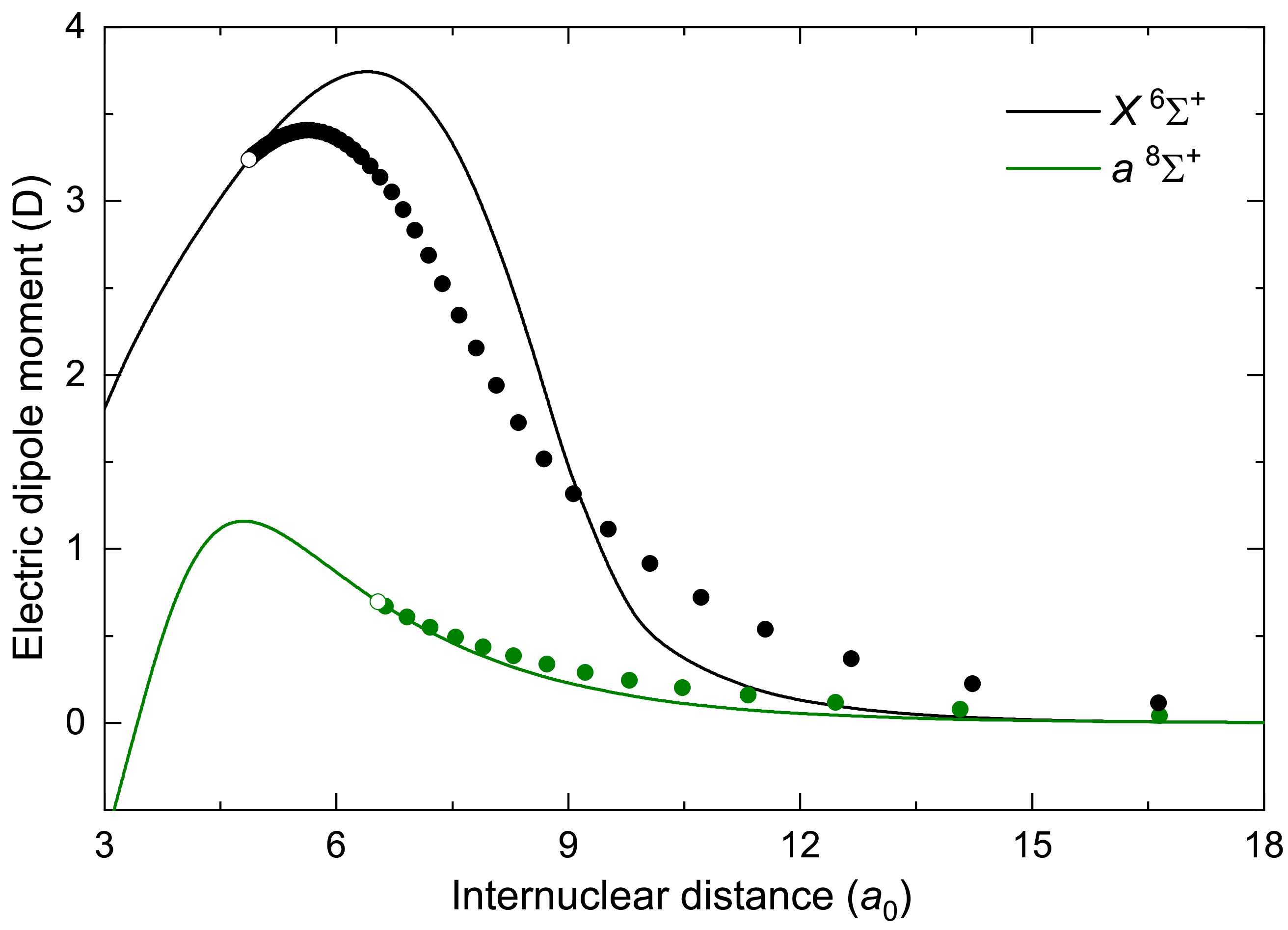}
\end{center}
\vskip -10pt
\caption{
Permanent electric dipole moments of the LiCr molecule in the $X\,^6\Sigma^+$ and $a\,^8\Sigma^+$ electronic states. The solid lines present them as a function of the internuclear distance. The empty dots mark their values at equilibrium distances, and the full dots present vibrationally averaged moments $d_v$ calculated for all vibrational levels $v$ supported by the considered electronic states placed at the corresponding vibrationally-averaged internuclear distances $R_v$ (two last points for both states are beyond the plot range).}
\label{fig:PEDM}
\end{figure}

We predict the well depths $D_e$ of the ground $X^6\Sigma^+$ and first-excited $a^8\Sigma^+$ states to be 8769(140)$\,$cm$^{-1}$ and 553(11)$\,$cm$^{-1}$ at the equilibrium distances $R_e$ of 4.87(1)$\,a_0$ and 6.50(2)$\,a_0$, respectively (see Fig.~\ref{fgr:OverviewMerged}(a)). As shown in Fig.~\ref{fig:PEDM}, a large permanent electric dipole moment of 3.3(2)$\,$D is expected for $^6$Li$^{53}$Cr in the rovibrational ground state ($v=0$), with even slightly larger values for excited vibrational levels of $X^6\Sigma^+$, up to 3.4(2)$\,$D for $v=14$. Additionally, we remark how sizable electric dipole moment values, exceeding 1$\,$D, are expected already for comparably high vibrational states ($v\lesssim37$) of the $X^6\Sigma^+$ potential, see Fig.~\ref{fig:PEDM}. At the same time, also the high-spin $a^8\Sigma^+$ state exhibits a significant permanent electric dipole moment of 0.67(3)$\,$D in its $v=0$ level (see Fig.~\ref{fig:PEDM}), a value exceeding the largest one found for  spin-triplet  bi-alkali molecules \cite{Tomza2023}.

For $^6$Li$^{53}$Cr, the ground electronic state supports $46(1)$ vibrational levels and the dissociation energy of the $v=0,j=0$ level is $D_{0}=8599(140)\,$cm$^{-1}$. The high-spin state supports exactly 16 vibrational levels with $D_{0}=512(11)$cm$^{-1}$. The rotational constants $B_0$ for $v=0$ levels of the $X\,^6\Sigma^+$ and $a\,^8\Sigma^+$ states are 14.02(4)$\,$GHz and 7.67(3)$\,$GHz. The bound vibrational levels exist up to $j_\text{max}=146(2)$ for $X^6\Sigma^+$ and $j_\text{max}=46(1)$ for $a^8\Sigma^+$. Precise knowledge of the number of vibrational states allows us, via mass scaling, to infer the scattering properties of all Li-Cr isotopic combinations from accurate experimental data obtained for $^6$Li-$^{53}$Cr~\cite{companion,PhysRevLett.129.093402}.

\subsection{Benchmarking the theory}

The high accuracy of our \textit{ab initio} calculation, combined with the comparably small values of both the Li-Cr reduced mass and well depth of the $a^8\Sigma^+$ state, allows us to obtain -- purely from first principles -- reliable predictions also for the scattering properties associated with this high-spin potential. These are encoded in the octet scattering length $a_8$, whose value of $41.48(2)\,a_0$ was determined with high precision in our Ref.~\cite{PhysRevLett.129.093402}, as the result of a global fit of a quantum collisional model to 52 experimentally observed FRs. 
In this work, our \textit{ab initio} model, without any  experimental input, predicts for the $^6$Li-$^{53}$Cr isotopic pair $a_8=49^{+48}_{-21}\,a_0$, which is in excellent agreement with the measured value. Our prediction results from a last vibrational level of the $a^8\Sigma^+$ state ($v\!=\!-1$) with  energy $E_{-1}=-735^{+640}_{-1600}\,$MHz and a van der Waals dispersion coefficient $C_6=954\,E_ha_0^6$ from perturbation theory \cite{companion}, both in good agreement with the fitted experimental values values of $E_{-1}=-1146\,$MHz and $C_6=922(6)\,E_ha_0^6$, respectively~\cite{PhysRevLett.129.093402}. Obtaining reliable predictions for the scattering properties of a complex, many-electron system, as the Li-Cr one, from first-principle calculations, is an extremely challenging task, and such a good agreement is thus remarkable. This makes our result the first-ever scattering length prediction for atom pairs involving a transition-metal element, and Li-Cr represents the second instance, after Li-Na~\cite{GronowskiPRA20}, where our \textit{ab initio} methods could quantitatively reproduce the scattering properties between two atomic species, both heavier than H, H$_2$ or He.

As an additional test, our \textit{ab initio} model reproduces correctly the order of magnitude of the subtle fine structure corrections to the spin-spin interactions resulting from the second-order spin-orbit coupling and screening direct magnetic dipolar interactions between electrons~\cite{companion}. The quantitative agreement between our \textit{ab initio} results and our data on ultracold Li-Cr collisions~\cite{PhysRevLett.129.093402} -- governed by the least-bound vibrational level and long-range tail of the interatomic potential -- lets us conclude that our model can reliably provide information, with good predictive power, also on other LiCr properties, so far experimentally unexplored.

\section{Optical manipulation of L\MakeLowercase{i}C\MakeLowercase{r}}
\label{Sec:TheorySTIRAP}
As anticipated at the beginning of Sec. \ref{Sec:Theory}, coherent two-photon optical transfer of LiCr Feshbach dimers to the rovibrational ground state requires detailed knowledge of the intermediate, electronically-excited  levels, of the associated transition dipole moments, and of the resulting molecular spectra --  completely lacking experimental investigation so far. In the following, we employ our \textit{ab initio} model to explore the properties of excited electronic states of LiCr, which allows us to identify suitable transitions  for the efficient conversion of Feshbach dimers into tightly-bound states, and more generally for the optical manipulation of LiCr.

\subsection{Excited electronic states}

\begin{figure}[t]
\begin{center}
\includegraphics[width=\linewidth]{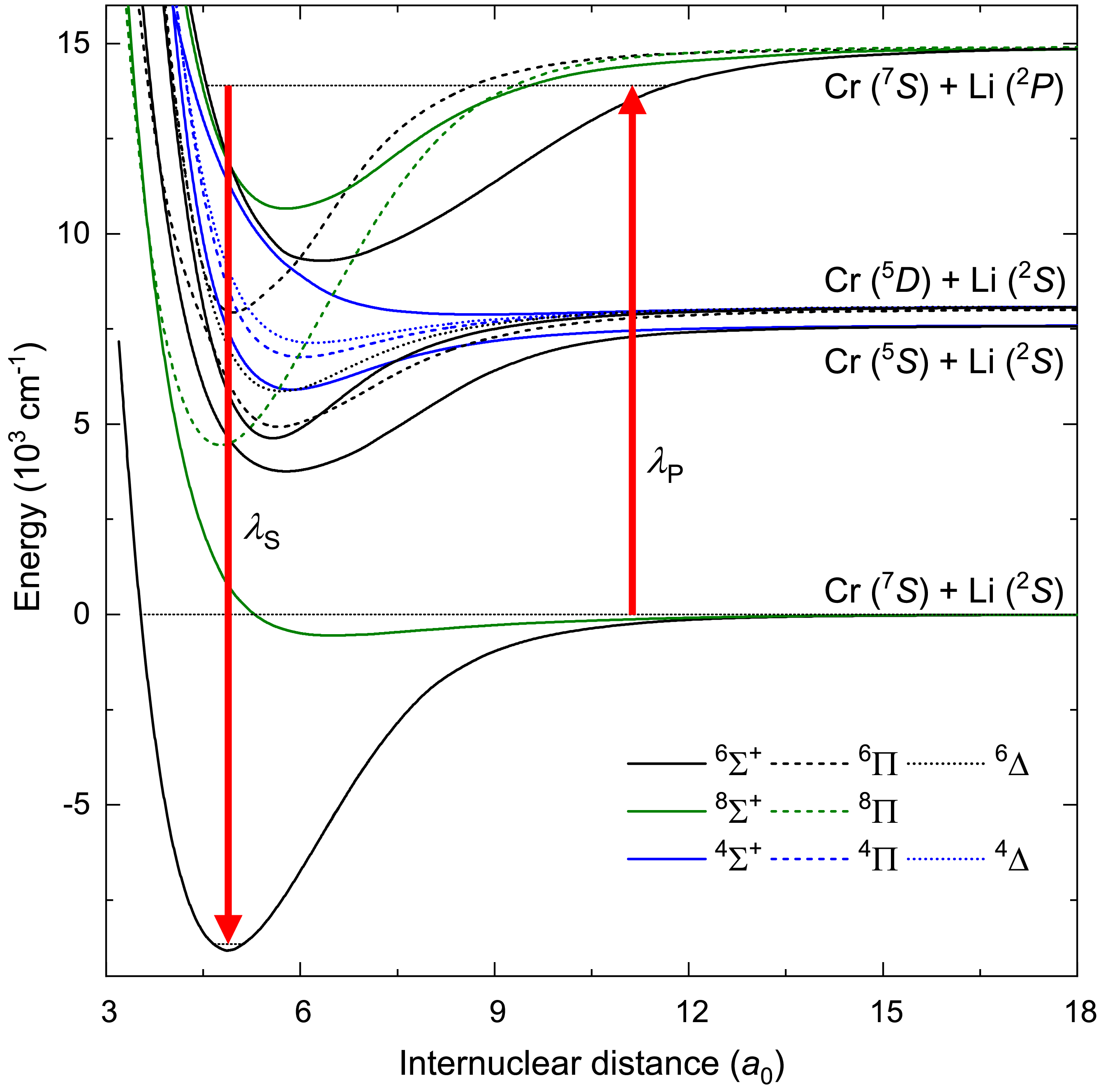}
\end{center}
\vskip -10pt
\caption{Potential energy curves of the LiCr molecule in the ground and excited electronic states (in the Hund's case (a) representation). The arrows show the proposed pump and Stokes STIRAP transitions at $\lambda_\text{P}\approx 720\,$nm and  $\lambda_\text{S}\approx 445\,$nm, respectively.}
\label{fig:PECs}
\end{figure}

Calculations of excited electronic states are significantly more challenging than those of ground ones. Excited states of the LiCr molecule are especially demanding because of the highly multireference nature of the Cr half-filled electron $d$-shell participating in chemical bonding. In Figure~\ref{fig:PECs}, we present the spectrum in the Hund's case (a) representation of all molecular excited electronic states of LiCr up to the Cr($^7S$)+Li($^2P$) asymptote. The states of $\Sigma$, $\Pi$, and $\Delta$ spatial symmetries with total electronic spin $S=3/2, 5/2, 7/2$ exist. The higher-lying energy asymptote, Cr($^7P$)+Li($^2S$), is above 20 000$\,$cm$^{-1}$, and it will not be discussed here. 

Two families of excited electronic states can be seen in Fig.~\ref{fig:PECs}: A first set dissociating into metastable, excited-state Cr ($^5S$ or $^5D$) and ground-state Li ($^2S$), and another set that dissociates into ground-state Cr ($^7S$) and excited-state Li ($^2P$). The former ones are relatively shallow, especially those associated with Cr in the $^5D$ state with the closed $4s^2$ shell screening interactions with the open $3d^4$ shell. The states belongnig to the second family, and connected to the Cr($^7S$)+Li($^2P$) asymptote, are deeper. Despite the large number of states, an avoided crossing is visible only between those with  $^6\Pi$ symmetry.

The spin-orbit coupling can couple the excited electronic states of different spin and spatial symmetries. Its largest strength is expected for states dissociating into Cr($^5D$)+Li($^2S$), which is asymptotically split by hundreds of cm$^{-1}$. Coupling of crossing states leads to avoided crossings and to enhanced mixing of different spin components. The spin-orbit coupling between electronic states dissociating into Cr($^7S$)+Li($^2P$) should be smaller because of its negligible asymptotic value, although it might be enhanced at small internuclear distances by the presence of Cr. Although this will be discussed in more detail in the following section, we remark here how the energy landscape of LiCr, shown in Fig.~\ref{fig:PECs}, is significantly richer than the more familiar one of bi-alkali systems, and it originates from the involvement of $d$-shell electrons of Cr. 
While this results in a more complex scenario, the existence of metastable excited states such as $^4\Delta$ -- with large angular momenta and potentially long lifetime  -- opens the way for applications of LiCr, primarily in the context of precision measurements~\cite{RoussyScience23,KondovNP2019}, unattainable with alkali-metal dimers.

\subsection{Prospects for STIRAP}

Having provided an overview of both ground- and electronically-excited states of LiCr  in the previous sections, we now move to discuss possible pathways for the efficient optical transfer of our Feshbach dimers towards deeply-bound levels via STIRAP~\cite{VitanovRMP17}. This method was successfully employed for a large number of alkali-metal diatomic molecules, and it involves two laser pulses that transfer coherently molecular population between the initial and final states through an intermediate electronically-excited state in a $\Lambda$ configuration, while never populating it. Accurate knowledge of the energies and dipole moments of the most favorable transitions are needed to execute STIRAP successfully. While such properties can be obtained experimentally through tedious and time-consuming spectroscopic measurements, identification of suitable STIRAP pathways can be highly accelerated when guided by state-of-the-art molecular calculations presented in this work. Efficient STIRAP transfer of Feshbach dimers  to the absolute ground state necessarily requires to pinpoint an intermediate, electronically-excited vibrational state that exhibits a significant overlap with both the initial (weakly-bound, long-ranged) and the final (deeply-bound, short-ranged) levels. 
In our search, we focus on STIRAP paths involving sextet electronic states only. Restricting our survey to this sub-class is motivated by the fact that our Feshbach dimers have (almost) pure sextet character, see Sec.~\ref{Sec:MoleculeProperties_MagneticMoment}, as the one of the LiCr rovibrational ground state. This greatly simplifies our task, and the corresponding STIRAP scheme is conceptually equivalent to  the singlet-to-singlet optical transfer of bi-alkali dimers, recently and successfully employed to produce ground-state $^6$Li$^{40}$K molecules~\cite{YangPRL20}.

\begin{figure}[tb!]
\begin{center}
\includegraphics[width=1\linewidth]{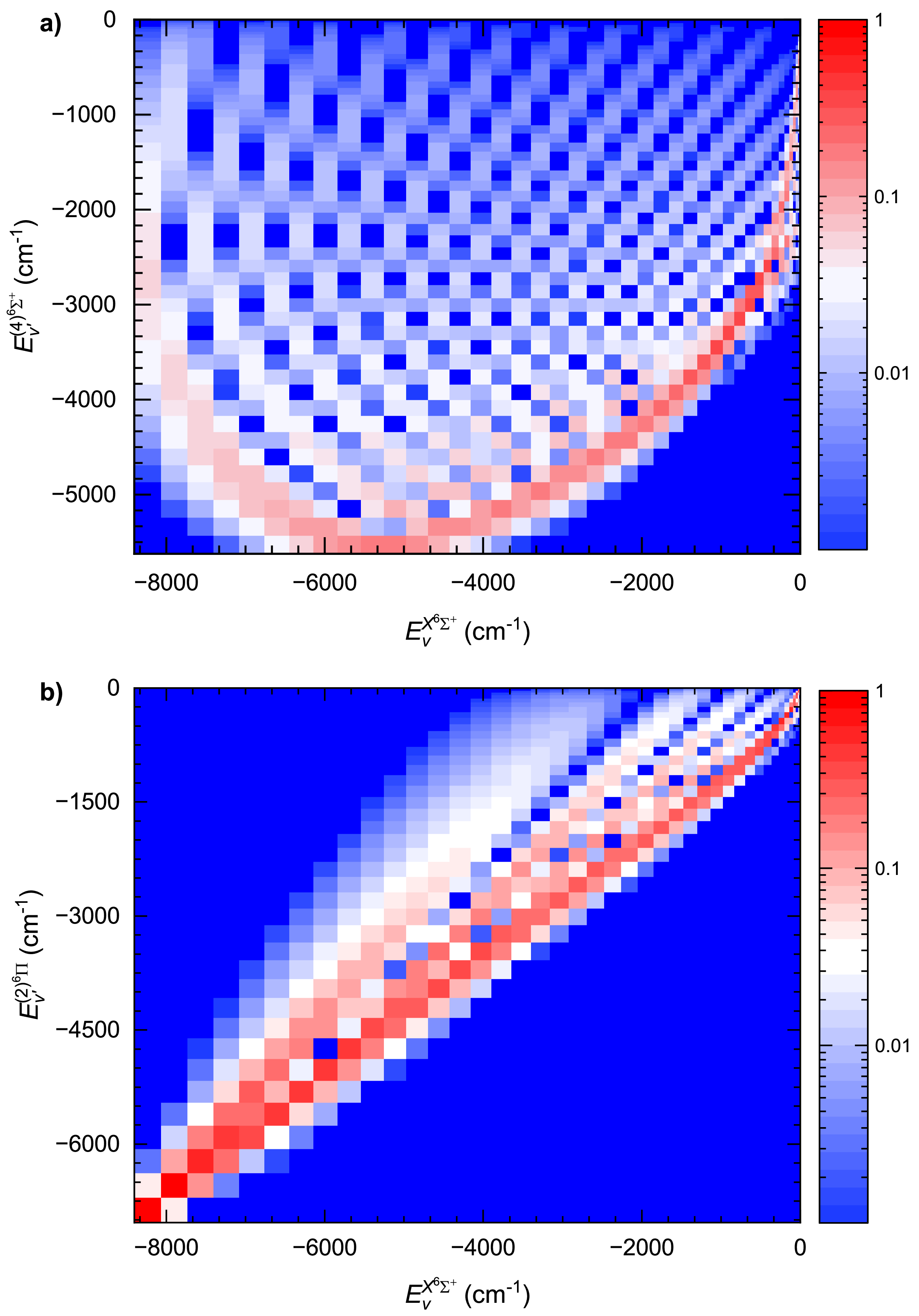}
\end{center}
\vskip -10pt
\caption{Franck-Condon factors $|\langle v|v'\rangle|^2$ between vibrational levels of the LiCr molecule in the $X\,^6\Sigma^+$ ground electronic state $v$ and $(4)\,^6\Sigma^+$ and $(2)\,^6\Pi$ excited electronic states $v'$ as a function of their energies $E_v$: (a) $X^6\Sigma^+\to (4)^6\Sigma^+$ and (b) $X^6\Sigma^+\to (2)^6\Pi$.}
\label{fig:FCFs}
\end{figure}

The $(4)^6\Sigma^+$ and $(2)^6\Pi$ electronic states dissociating into Cr($^7S$)+Li($^2P$) are the most promising for providing suitable intermediate levels because they are relatively well separated from other states and should be accessible at convenient optical wavelengths by strong transition dipole moments borrowed from the strong atomic transition ${^2S}\to{^2P}$ in Li~\cite{companion}. According to our calculations, they support 54(5) and 48(4) vibrational levels, respectively, while their dissociation energies are around 5700$\,$cm$^{-1}$ and 6800$\,$cm$^{-1}$. Franck-Condon factors (FCFs), $|\langle v|v'\rangle|^2$, between vibrational levels supported by the ground ($v$) and excited  ($v'$) electronic states give the initial insight into possible optical transitions. They are presented in Fig.~\ref{fig:FCFs} for the relevant sextet states. The overview of the FCFs suggests the $X\,^6\Sigma^+\to(4)\,^6\Sigma^+$ transitions (see Fig.~\ref{fig:FCFs}(a)) as the most promising ones for the STIRAP implementation. Similar to alkali-metal dimers, the characteristic bent shape of the largest FCFs shows the existence of intermediate levels of the excited $(4)\,^6\Sigma^+$ state having noticeable overlap with both weakly- and deeply-bound levels of the ground $X\,^6\Sigma^+$  state. In contrast, the FCFs for the $X\,^6\Sigma^+\to(2)\,^6\Pi$ transitions (see Fig.~\ref{fig:FCFs}(b)) are visibly diagonal, which is not a preferable pattern for an efficient STIRAP transfer.

\begin{figure}[tb!]
\begin{center}
\includegraphics[width=1\linewidth]{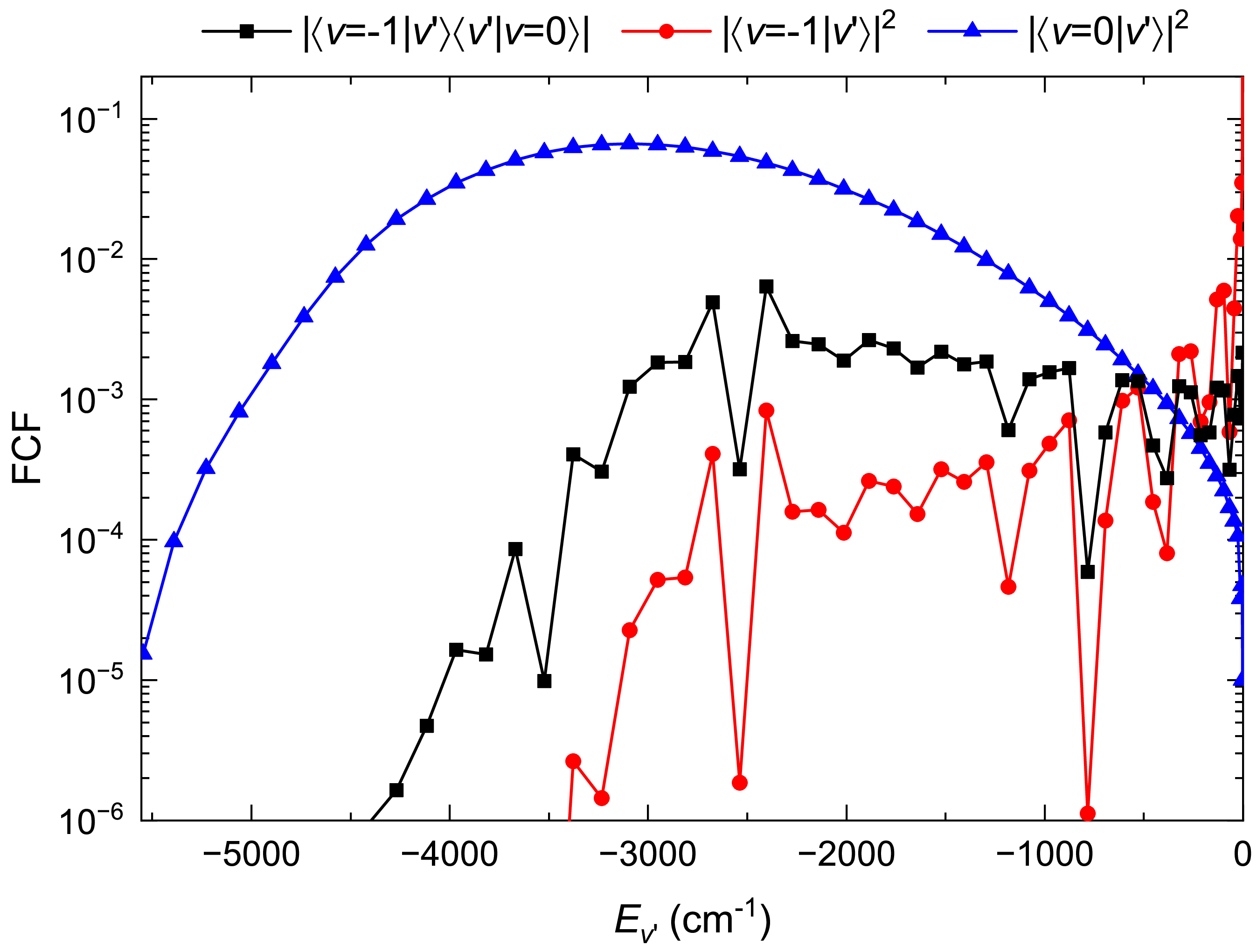}
\end{center}
\vskip -10pt
\caption{Franck-Condon factors $|\langle v|v'\rangle|^2$ between the initial weakly-bound $v=-1$ and final ground $v=0$ vibrational levels of the $X\,^6\Sigma^+$ ground electronic state and the intermediate vibrational levels $v'$ with energies $E_v'$ of the $(4)\,^6\Sigma^+$ excited electronic state and their combination $\sqrt{|\langle -1|v'\rangle|^2|\langle 0|v'\rangle|^2}=|\langle -1|v'\rangle\langle v'|0\rangle|$ for two-photon transition between $v=-1$ and $v=0$ levels.}
\label{fig:TMs}
\end{figure}

Fig.~\ref{fig:TMs} presents the FCFs for transitions from the least-bound (denoted here as $v=-1$) and ground ($v=0$) levels of the $X\,^6\Sigma^+$ ground electronic state to intermediate levels $v'$ supported by the $(4)\,^6\Sigma^+$ excited electronic state. The largest FCFs for the ``pump" transition $v=-1\to v'$ (red symbols) are expected to connect to the most weakly-bound excited-state levels just below the Cr($^7S$)+Li($^2P$) atomic threshold, but values above $10^{-4}$ are predicted up to binding energies $E_{v'}$ of almost 3000$\,$cm$^{-1}$. These FCFs also give the shape of a photoassociation spectrum. 
The largest FCFs for the ``Stokes" transition $v'\to v=0$ (blue symbols) are found to connect to levels in the middle of the interaction potential well, with values above $10^{-2}$ for binding energies between 1400$\,$cm$^{-1}$ and 4400$\,$cm$^{-1}$. Strong Stokes transitions result from the alignment of the equilibrium distance of the $X\,^6\Sigma^+$ ground electronic state and the inner classical turning point  of the $(4)\,^6\Sigma^+$ excited electronic state (see Fig.~\ref{fig:PECs}). The combination of the pump and Stokes FCFs govern the STIRAP. The largest two-photon FCF, $|\langle -1|v'\rangle\langle v'|0\rangle|$, of $6\cdot 10^{-3}$, is found for the $v'=20$ intermediate level with a binding energy around 2400$^{-1}$. Yet, FCFs exceeding $10^{-3}$ are expected for most vibrational levels with binding energies up to 3100$\,$cm$^{-1}$. Thus, to avoid a large power imbalance of the pump and Stokes laser fields, intermediate $v'$ levels with smaller binding energies around 1000$\,$cm$^{-1}$ can be addressed with convenient wavelengths of $\lambda_\text{P}\approx 720\,$nm and $\lambda_\text{S}\approx 445\,$nm, respectively, see Fig.~\ref{fig:PECs}. 

The predicted FCFs for the $X\,^6\Sigma^+\to(4)\,^6\Sigma^+$ transitions are similar to, or even more favorable than, those found for alkali-metal diatomic molecules, where a STIRAP transfer efficiency as large as 90$\,$\% was achieved. Our predictions are robust against the uncertainty of calculated excited electronic states~\cite{companion}, and favorable conditions are persistent, assuming 10-20$\,$\% errors of the well depth and 0.2-0.5$\,a_0$ errors or the equilibrium distance. Including the $R$-dependence of the transition dipole moments also does not affect the presented findings~\cite{companion}. Also spin-orbit coupling -- neglected in these calculations and that could perturb some levels accidentally close to other ones of different symmetry -- is expected, overall, to negligibly affect our results.

The detailed analysis of the $X\,^6\Sigma^+\to(2)\,^6\Pi$ transitions, see Fig.~\ref{fig:FCFs}(b), confirms that they are less suited for STIRAP, as they exhibit two-photon FCFs smaller by several orders of magnitude than those for $X\,^6\Sigma^+\to(4)\,^6\Sigma^+$ transitions~\cite{companion}. The pump FCFs are smaller because the $(2)\,^6\Pi$ potential is shallower than the $(4)\,^6\Sigma^+$ one at intermediate and large distances (see Fig.~\ref{fig:PECs}). The FCFs noticeably exhibit a diagonal structure, that results from the alignment of the equilibrium distances of the $X\,^6\Sigma^+$ and $(2)\,^6\Pi$ states. This peculiar feature, inconvenient for STIRAP, may instead allow for direct imaging, and possibly even laser cooling, of ground-state LiCr molecules. While the exact values of diagonal FCFs are very sensitive to the accuracy of underlying electronic states, our model yields FCFs  robustly above 0.5~\cite{companion}, sufficiently large to enable direct optical imaging of ground-state LiCr, e.g.,~in optical tweezers. Additionally, for our current PECs, the FCFs for $v'=0\to v=0$, $v'=0\to v=1$, and $v'=0\to v=2$ transitions are predicted to be 0.94, 0.056, and 0.0007, respectively. 
Such a set of values is comparable to those of CaF and similar molecules and, if experimentally confirmed, could even allow for direct laser cooling of LiCr.

\section{Conclusions and outlook}
\label{Sec:Conclusions}
In summary, we have produced ultracold gases of up to $50 \times 10^3$ bosonic $^6$Li$^{53}$Cr Feshbach molecules from atomic Fermi mixtures, reaching phase-space densities of 0.1 at about 200\,nK. Thanks to the immunity to two-body decay and the good stability against three-body recombination of our mixture, we could perform magneto-association with $B$-field rates slower than the two-body adiabatic regime by orders of magnitude. We have directly revealed the paramagnetic nature of the LiCr electronic ground-state, and demonstrated precise control of the Feshbach state via a novel optical measurement of the open-channel fraction and binding energy. Through the characterization of the relevant loss mechanisms affecting Feshbach dimers, we have identified an experimental configuration where their lifetime exceeds 0.2\,s. By developing a quantum-chemical model for LiCr via state-of-the-art \textit{ab initio} methods, we have determined the fundamental properties of this new molecular species. In particular, for the rovibrational $X^6\Sigma^+$ ground state we predict a large electric dipole moment of 3.3(2)\,D, on top of a sizable magnetic one (5\,$\mu_\textrm{B}$). Additionally, our model foresees that Feshbach dimers, already created in the least-bound $X^6\Sigma^+$ vibrational level, can be efficiently transferred, via STIRAP through a $(4)^6\Sigma^+$ level, to the absolute ground state. There, direct imaging and cooling schemes may be enabled by $X^6\Sigma^+\rightarrow(2)^6\Pi$ optical transitions. 
Our results thus make ultracold LiCr emerge as an appealing system for a wealth of fundamental studies and future applications. 

A high phase-space density sample of $^6$Li$^{53}$Cr Feshbach dimers, already realized, opens exciting new routes for the investigation of strongly-correlated fermionic matter. Extending the protocols developed in this work may allow us to Bose-condense Feshbach dimers -- thereby paving the way to BCS-BEC crossover studies in presence of a large mass asymmetry \cite{Gezerlis2009,Baarsma2010,Pini2021}. The creation of long-lived LiCr dimers in the presence of a controlled amount of Cr  is a fundamental step towards novel few- and many-body phenomena, uniquely enabled by the peculiar mass ratio of our atomic Fermi mixture \cite{VarennaNotesZaccanti2022}: For instance, the existence of LiCr$_2$ fermionic trimers \cite{Kartavtsev2007, Levinsen2009,Endo2011}, and LiCr$_3$ bosonic tetramers \cite{Blume2012A,Bazak2017, Liu2022A}, the emergence of new types of quasi-particles within the light-impurity problem \cite{Mathy2011,Liu2022B} and, possibly, the appearance of many-body regimes beyond the BCS-BEC crossover scenario, such as trimer Fermi gases \cite{Endo2016} or quartet superfluid states \cite{Liu2023}.

Our joint experimental and theoretical results suggest that 
the realization of quantum gases of doubly-polar $^6$Li$^{53}$Cr bosonic molecules, with large electric and magnetic dipole moments, is within reach. Clearly, identification of the optimal STIRAP transfer will require extensive laser spectroscopy, but this will be greatly facilitated by our \emph{ab initio} model predictions and by the long lifetime of Feshbach dimers, already achieved. Interestingly, electric dipole moments as high as 1\,D can be already obtained in relatively shallow vibrational levels of LiCr with $\nu\simeq 37$ and binding energy as low as 190\,cm$^{-1}$, see Sec.~\ref{Sec:Theory}. Given the relatively simple spectroscopic survey needed to find these states, this possibility represents an appealing intermediate, short-term step. In particular, such vibrationally-excited molecules offer high sensitivity to the electron-to-proton mass ratio $m_e/m_p$ \cite{companion}, overcoming that of alkali and alkaline-earth dimers, and they could be employed to detect possible variations of $m_e/m_p$ in precision measurements, and to gain insight into fundamental physics.

LiCr molecules, pinned in optical lattice sites or tweezer traps, will find immediate application in the context of quantum simulation of spin Hamiltonians \cite{Micheli2006ToolboxPolMol,Pérez-Ríos_2010,Yao2018,CornishRev}, as well as high-dimensional quantum computing \cite{PhysRevA.101.062308,10.1063/1.4942928}, exploiting the internal spin degree of freedom, absent in ground-state alkali dimers. Bulk gases will be ideal test-beds for quantum-controlled chemistry: Although ground-state LiCr molecules are chemically unstable under atom-exchange reactions LiCr+LiCr$\rightarrow$Li$_2$+Cr$_2$, in their spin-stretched state resulting from STIRAP they are expected to be stabilized because the decay to deep Cr$_2$ levels becomes spin-forbidden. The predicted rotational constant and electric dipole moment, incidentally close to those of CaF, suggest that collision shielding, via micro-wave \cite{doi:10.1126/science.abg9502} or static electric fields \cite{Li2021,PhysRevResearch.5.033097}, could be applied to stabilize bulk samples and perform evaporative cooling. Additionally, resonant tuning of magnetic \cite{Park2023} and field-linked resonances \cite{Chen2023} may be investigated in LiCr. Prospects for precision measurements also appear promising: on top of the measurement of $m_e/m_p$ already mentioned, the observation of the electron electric dipole moment may be facilitated by the internal co-magnetometer states offered by long-lived $^4\Delta$ metastables levels \cite{RoussyScience23}, unavailable in alkali-metal dimers. 

Finally, it is worth noting that fermionic isotopologues $^7$Li$^{53}$Cr and $^6$Li$^{52}$Cr could also be formed via magneto-association across suitable FRs \cite{PhysRevLett.129.093402} on our experimental setup. More generally, combining Cr with other alkalis, with heavier mass and lower electronegativity than Li, could result in molecular paramagnetic species with even larger electric dipole moments.

\section{Acknowledgements}
\label{Sec:Acknowledgements}
We thank A. Canali, R. Grimm, S. Meek, D. Petrov, G. Santambrogio, and the LENS Quantum Gases group for fruitful discussions, and G. Rosi and L. Salvi for technical support. We acknowledge A. Canali also for contributions to the experiment, and D. Petrov for sharing unpublished theoretical results. 
K.Z.-K., M. G. and M.T.~acknowledge Poland’s high-performance computing infrastructure PLGrid (HPC Center: ACK Cyfronet AGH) for providing computer facilities and support (computational grant no.~PLG/2023/016115).

This work was supported by the European Research Council under Grant No.~637738, by the EU H2020 Marie Skłodowska-Curie under Grant No.~894442 (CriLiN, fellowship to A.C.), by the National Science Center, Poland under Grants No.~2019/35/N/ST4/04504 and 2020/38/E/ST2/00564, by the Italian Ministry of University and Research under the PRIN2022 project No.~20227F5W4N and, co-funded
by the European Union—NextGenerationEU, under the ‘Integrated
infrastructure initiative in Photonic and Quantum Sciences’
 I-PHOQS, and under the Young Researcher Grants MSCA\_0000042 (PoPaMol, fellowship to A.C.) and  MSCA\_0000048 (MajorSuperQ, fellowship to M. S.). 

\appendix
\section{Trap Configuration and sample conditions}
\label{Appendix:TrapConfig}
The BODT used in this work is realized by superimposing an infrared laser beam with a green one, both propagating along the $x$ direction in the horizontal $(x,y)$ plane \cite{PhysRevA.106.053318}. The former is delivered by a multimode 1070\,nm fiber laser (IPG Photonics YLR-300) and has a horizontal (vertical) waist of $w_y=58 \,$\textmu m ($w_z=44 \,$\textmu m). The latter is generated by a single-mode source at 532\,nm (Lighthouse Photonics Sprout-G) and has horizontal (vertical) waist of $w_y=48 \,$\textmu m ($w_z=45 \,$\textmu m). The 1560\,nm FORT employed to obtain long-lived molecular samples, see Sec.~\ref{Sec:MoleculeLifetime}, has a circular beam waist of $55\,$\textmu m, and it is delivered by an Erbium fiber amplifier (Keopsys-CEFA-C-BO-HP) seeded by a single-mode master laser (Rock™, NP Photonics).

Our magnetic field coils provide, besides the required bias, both a magnetic field curvature
in the $(x,y)$ horizontal plane of 12.3\,G cm$^{-2}$ -- which determines the axial confinement along the weak axis of the BODT -- as well as a magnetic field gradient, of about 1.5\,G cm$^{-1}$ along the vertical direction. The latter value is optimized in order to levitate the heavier Cr atoms -- that would otherwise be lost from the trap at the end of the evaporation stage -- and, simultaneously, to minimize the gravitational sag between the two mixture components.

Typical trap frequencies for most of the experimental studies described in the main text,
are $(\nu_x, \nu_y,\nu_z)\!=\!(17,115,156)\,$Hz and $(\nu_x, \nu_y,\nu_z)\!=\!(14,124,118)\,$Hz for Li and Cr, respectively, with an overall trap depth ratio of $U_{\textrm{Li}}/U_{\textrm{Cr}}\!=\!0.75$, which can be tuned thanks to the species-dependent dynamic polarizabilities \cite{AtomicPolarizabilities} of the two BODT beams.

 To optimize magneto-association, we maximize the spatial density overlap between Li and Cr clouds by tuning the power ratio of the BODT. We do this by linearly ramping the green ODT from the final evaporation value, which provides a deeper trap for Cr with respect to Li, to a lower value right before the magneto-association field sweep. During this ramp the trap gets tighter and deeper for Li, while the opposite holds for Cr. A sweet spot exists for a green to IR power ratio of about 0.35, which corresponds to $P_G=60\,$mW and $P_{IR}=180\,$mW, where $\langle n_{\textrm{Li}} n_{\textrm{Cr}} \rangle=5.0(5)\times10^{16}\,$cm$^{-3}$. The confinement in this BODT configuration is characterized by geometrically averaged radial trapping frequencies $\sqrt{\nu_y \nu_z}$ of about 340\,Hz, 105\,Hz and 145\,Hz for Li, Cr and LiCr dimers, respectively. The   axial confinement, as already mentioned above, is entirely set by our magnetic-field curvature, and it is characterized by  trapping frequencies of $\nu_{x,\textrm{Li}}= 17.0(6)$Hz, $\nu_{x,\textrm{Cr}}= 14.0(5)$Hz, and $\nu_{x,\textrm{LiCr}}= 12.1(5)$Hz,
 where LiCr are assumed to be fully closed-channel molecules. After the switch-off of the BODT, atoms and molecules expand into a magnetic saddle potential, with trapping frequencies $\nu_x=\nu_y$ in the horizontal plane --equal to the \emph{in situ} axial ones-- and vertical anti-trapping frequencies of 
 $\nu_z= i \sqrt{2}\,\nu_x$.
 
\section{Magnetic-field stability}
\label{Appendix:FieldStabilization}

\begin{figure}[t]
\centering
  \includegraphics[width=0.9\columnwidth]{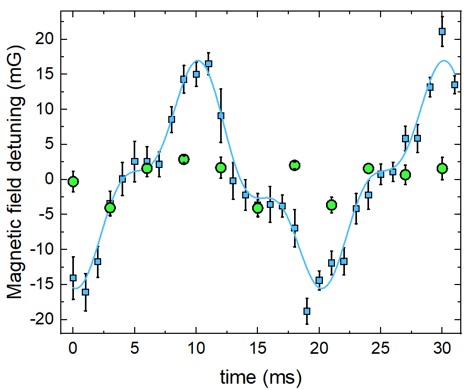}
  \caption{Stability of the bias magnetic field at 1.4\,kG. Blue squares show a typical magnetic-field trace, obtained through the spectroscopic technique described in the text after synchronizing the experimental sequence with the 50\,Hz mains, once the sole active stabilization of the current of the FESH1 and FESH2 coils setups is implemented. The observed trace, synchronous with the mains and characterized by a peak-to-peak amplitude of about 40\,mG, is well fitted to a sum of sinusoidal functions with a main Fourier component at 50\,Hz plus a few higher harmonics of decreasing amplitude (blue line). Employing the (reversed) fitted trace to drive the COMP coils setup, we remove the main (synchronous) noise contribution, being left with only a residual asynchronous noise (green circles), characterized by a standard deviation of about 2.5\,mG.}
  \label{fgr:MagFieldStab}
\end{figure}
The experimental studies described in the main text require an excellent stability of the magnetic-field bias, and an accurate dynamic tuning of it within the experimental cycle, in light of the comparatively narrow width of the available Li-Cr FRs combined with their high-field location, exceeding 1.4\,kG. 
In our setup, a highly-stable magnetic field is produced by means of three different sets of coils, denoted as FESH1, FESH2, and COMP coils, respectively. 
The main FESH1 coils are able to produce  magnetic fields of up to 1.5\,kG, corresponding to a DC current of about 200\,A. This is generated by a 15\,kW  programmable power supply (TDK-Lambda GSP-80-195). The FESH1 current, measured with a current transducer (LEM ITN 600-S ULTRASTAB), is actively stabilized via a proportional-integral controller acting on the voltage programming input of the power supply.
This ensures reproducibility and medium-term stability of the applied bias field. 
The (much smaller) FESH2 coils can produce fields of about 2.5\,G at the maximum employed current of 5\,A. They are used, in combination with FESH1, to finely tune the magnetic field around the FR. The FESH2 current is as well actively stabilized by means of a similar setup as the FESH1 current with a second set of controller and current transducer of the same type.
As the FESH2 setup is concerned, active stabilization of the current allows us to reduce the noise of the generated magnetic field well below 1\,mG. 

The FESH1 coils, instead, that are driven with a significantly higher current, exhibit a larger residual closed-loop noise synchronous with the 50\,Hz mains. This is characterized by a peak-to-peak amplitude of about 40\,mG at 1.4\,kG and by Fourier components at 50\,Hz and a few higher harmonics, see blue squares and blue curve in Fig.~\ref{fgr:MagFieldStab}.
This behavior is revealed by synchronizing the experimental sequence with the 50 Hz mains. Since the line carrier frequency can vary appreciably within the duration of the experiment, we not only synchronize the start of the experiment with the AC line but also before the execution of field-sensitive measurements. This allows us to trace the magnetic field fluctuations at $t\geq t_1$ via RF-spectroscopy on the $\textrm{Cr}|1\rangle \leftrightarrow \textrm{Cr}|2\rangle$ Zeeman transition. Specifically, we adjust the RF power and pulse duration to achieve a $\pi$-pulse on resonance, and then detune the frequency to yield a 50\% transferred fraction. This side-of-fringe configuration yields the maximum sensitivity to magnetic-field variations and, for Fourier-limited pulse widths exceeding noise-induced frequency shifts, it can be used to unambiguously retrieve the time evolution from the transfer efficiency. For this reason, we initially employ a 1.8\,ms-long pulse to characterize the noise of our coils setup, and only at a later stage we extend it to 3.6\,ms to gain in sensitivity at the expense of a reduced bandwidth. A typical magnetic-field trace obtained with the initial 1.8\,ms-long pulse is presented in Fig.~\ref{fgr:MagFieldStab} (blue squares). As anticipated, the observed trend is well fitted by a sum of sinusoidal functions with a main 50\,Hz component and a few higher harmonics of progressively smaller amplitudes, see solid blue line. The best-fit function to the experimental data is then reversed and used to drive the additional COMP coils setup to implement the feed-forward to compensate field fluctuations. A second iteration with the 3.6\,ms-long pulse allows us to further refine the signal sent to the COMP coils. Through this active compensation procedure, we are finally left with residual 5\,mG peak-to-peak fluctuations synchronous with the 50\,Hz mains, see green circles in Fig.~\ref{fgr:MagFieldStab}. Such a stability is found to be maintained over several weeks.

In order to determine the shot-to-shot noise of the magnetic field, we acquire statistics of the observed transfer efficiency for variable spectral resolution at fixed time of the pulse $t_1$ during the sequence. The standard deviation of the samples is linearly correlated with the spectral resolution over a wide range and can be interpreted as shot-to-shot standard deviation of about $2.5\,$mG, which constitutes the main source for field fluctuations asynchronous with the 50\,Hz mains.
Finally, slow thermal effects on the whole coils setup cause a linear drift of the bias field of about 4\,mG/hour, which we compensate by finely adjusting the FESH2 coils current.

\section{Elastic collisions}
\label{Appendix:CollRates}
\subsection{The coupled oscillator model} 
The center-of-mass (COM) dynamics of the lithium and chromium cloud, explored throughout different interaction regimes as described in Sec.~\ref{Sec:FermiMixAndProcedures}, is analyzed in terms of a well-established model of two coupled oscillators, already tested on both homo- and hetero-nuclear atomic mixtures~\cite{PhysRevA.60.4734,PhysRevLett.87.173201,PhysRevLett.89.053202,FFerlaino_2003}. In this framework, the COM evolution of the two clouds along the axial ($x$) direction is described by two coupled differential equations that we express as:
\begin{align}
 m_{\mathrm{Cr}} \ddot{x}_{\mathrm{Cr}}&=-m_{\mathrm{Cr}} \omega_{\mathrm{Cr}}^2 x_{\mathrm{Cr}}- \frac{ 4m}{3} \gamma_\mathrm{el}\left(\dot{x}_{\mathrm{Cr}}-\dot{x}_{\mathrm{Li}}\right)&\nonumber \\ 
  m_{\mathrm{Li}} \ddot{x}_{\mathrm{Li}} &=-m_{\mathrm{Li}} \omega_{\mathrm{Li}}^2 x_{\mathrm{Li}}-  \frac{ 4m N_{\mathrm{Cr}}}{3N_\mathrm{Li}} \gamma_\mathrm{el}\left(\dot{x}_{\mathrm{Li}}-\dot{x}_{\mathrm{Cr}}\right).& \label{eq_coupled_osc}
\end{align}
Here $\omega_{\mathrm{Cr(Li)}}$ and $N_{\mathrm{Cr(Li)}}$ denote the Cr (Li) axial trap frequency and atom number, respectively, and $m
=m_{\mathrm{Li}}m_{\mathrm{Cr}}/(m_{\mathrm{Li}}+m_{\mathrm{Cr}})$ is  the Li-Cr reduced mass.  The sloshing dynamics of the two clouds are coupled through the right-most damping terms in Eqs.~(\ref{eq_coupled_osc}). The damping rates scale with the mean number of  elastic Li-Cr collisions  per unit of time  experienced by each component, and are here expressed in terms of the scattering rate per minority Cr atom $\gamma_\mathrm{el}$.  By fixing the axial frequencies and initial conditions ($x_{\mathrm{i}}(0), \dot{x}_{\mathrm{i}}(0)$) for $\mathrm{i}=\mathrm{Li}, \mathrm{Cr}$ to their experimentally-determined values, and accounting for the measured atom number evolution, we fit Eq.~(\ref{eq_coupled_osc}) to the recorded oscillation dynamics -- see examples in the right panels of Fig.~\ref{fig_hydro}(b) -- and thus extract $\gamma_\mathrm{el}$ as the sole free fitting parameter. 
The experimentally-determined Cr collision rate is compared in Fig.~\ref{fig_hydro}(c) with the theory expectation~\cite{Mosk2001} that we evaluate as:
\begin{equation}
    \gamma_\mathrm{el} = \langle n_\mathrm{Li}\rangle_\mathrm{Cr} \langle \sigma(a,R^*,k) \times v\rangle_T.
    \label{gammaEL}
\end{equation}
Here $\langle n_\mathrm{Li}\rangle_\mathrm{Cr}=\int n_\mathrm{Li}(\textbf{r}) n_\mathrm{Cr}(\textbf{r}) d\textbf{r}/N_\mathrm{Cr}$ denotes the  lithium density averaged over the chromium one, $\hbar k$ is the relative momentum for Li-Cr collisions and $v=\hbar k/m$ is the corresponding relative velocity. $\sigma(a,R^*,k)$ represents the (field- and momentum-dependent) elastic cross section, and $\langle \ldots  \rangle_T$ indicates thermal averaging over all relative momenta.
To a good approximation, $\langle n_\mathrm{Li}\rangle_\mathrm{Cr}$
remains constant to the experimentally-determined value of $\unit[1.2\times 10^{11}]{cm^{-3}}$ at all detunings and evolution times, given the moderate atom loss of the Li majority component. 
The elastic cross-section is given by (see e.g. Ref. ~\cite{VarennaNotesZaccanti2022})
\begin{equation}
 \sigma(a,R^*,k)=\frac{4 \pi a^2}{\left(1+(k a)^2 \frac{R^*}{a}\right)^2+(k a)^2},
 \label{Eq:CrossSection}
\end{equation}
where $R^*=6000\,a_0$ and $a(\delta B)=a_{bg} (1-\Delta B/\delta B)$, with $a_{bg}=41.48\,a_0$ and $\Delta B=0.48\,$G for our FRs. 
For simplicity of the calculation, thermal averaging of $\langle \sigma(a,R^*,k)  k\rangle_T $ is performed by neglecting the Cr thermal distribution within the momentum integral. This is justified by the fact that, owing to the large mass imbalance of our mixture, $m_\mathrm{Cr}/m_\mathrm{Li}\simeq 8.8$, the Li-Cr relative velocity is essentially set by the sole lithium one. Under this approximation, averaging over a Fermi-Dirac distribution at $T/T_F\sim\!0.3$, or over a Boltzmann-gas weight at an effective temperature of $T= \unit[700]{nK}$, yields nearly identical results throughout the interaction regime we explored. We employ Eq.~(\ref{gammaEL}) to fit the experimental data, with the resonance pole position $B_0$ as single free parameter implicitly entering Eq.~(\ref{Eq:CrossSection}), while fixing all other quantities to the corresponding values experimentally determined or given by the quantum-collisional model we developed in Ref. \cite{PhysRevLett.129.093402}.

\subsection{Transition from collisionless to collisional hydrodynamics regime}\label{sec_damping_rate}

Previous studies~\cite{PhysRevA.60.4734,PhysRevLett.87.173201,PhysRevLett.89.053202,FFerlaino_2003}, identified the transition from the collisionless to the hydrodynamic regimes by analyzing the scaling of the damping rate $\gamma_{\textrm{damp}}$ of the coupled oscillations with respect to the collision rate $\gamma_\mathrm{el}$. Specifically, in the collisionless regime $\gamma_{\textrm{damp}}$ linearly grows with $\gamma_\mathrm{el}$, whereas in the hydrodynamic regime it exhibits a progressive reduction proportional to $1/\gamma_\mathrm{el}$.

\begin{figure}[t]
    \centering
    \includegraphics[width=1\linewidth]{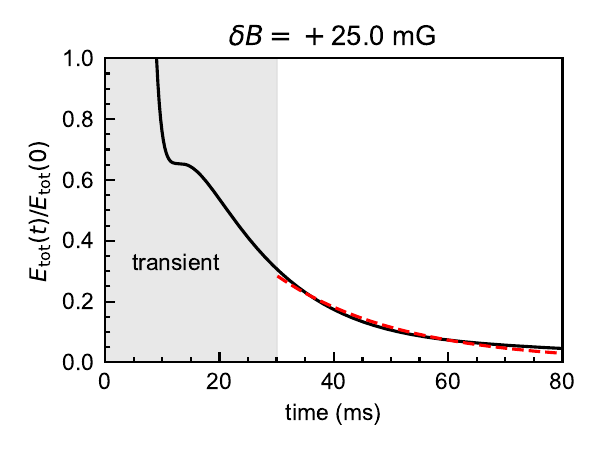}
\vskip -10pt
    \caption{Time evolution of the total energy, extracted from the fit of Eq.~(\ref{eq_coupled_osc}) to the data, characterized by an initial transient behaviour, and a subsequent almost pure exponential decay. Here, $E_{\textrm{tot}}$ is shown for $\delta B = \unit[25]{mG}$ (black solid line), corresponding to $\gamma_\mathrm{el}= \unit[490]{Hz}$. The red dashed line is the fitted exponential decay after the transient behaviour.}
    \label{fig:transient}
\end{figure}

Extracting $\gamma_{\textrm{damp}}$ from a fit to the COM dynamics is rather challenging, owing to the concurrent oscillatory motion, especially when the initial short-time evolution is concerned near the resonantly-interacting region, see the example of Fig.~\ref{fig_hydro}(b) at $\delta B=30\,$mG.

\begin{figure*}[t]
    \centering
    \includegraphics[width=0.8 \textwidth]{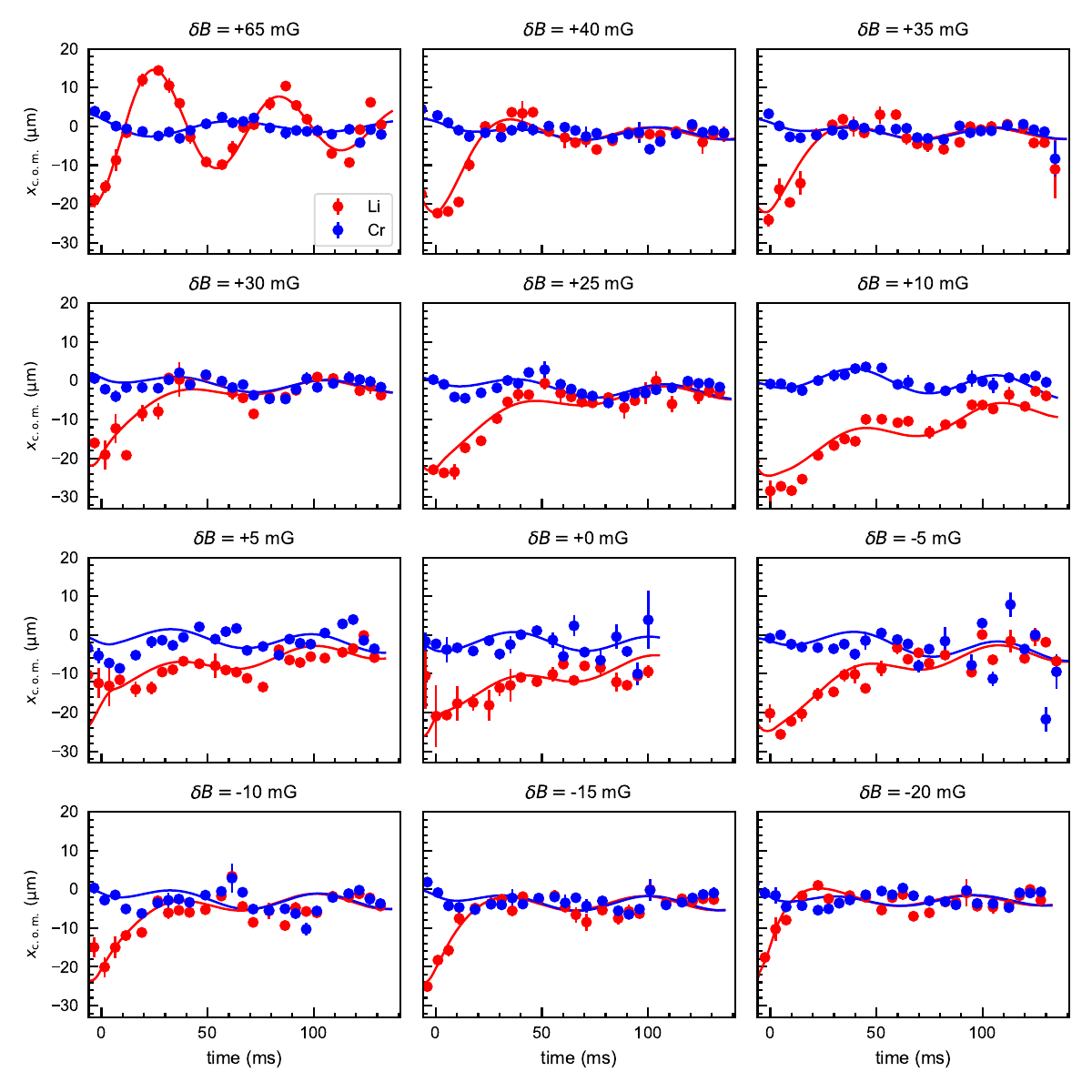}
    \vskip -10pt
    \caption{Examples of in-trap oscillations of Li and Cr COMs, recorded for different detuning $\delta B$ across the FR region, together with the fitted coupled oscillator model Eqs.~(\ref{eq_coupled_osc}) (solid lines).}
    \label{fig_osc_together}
\end{figure*}

A convenient way to extract the damping rates at all detunings, given  our limited observation time over a few oscillation periods, is to employ the fitted $ x_{\mathrm{Li}}(t) $ and $ x_{\mathrm{Cr}}(t) $ to obtain the total energy $ E_{\mathrm{tot}}(t) = E_{\mathrm{kin}}(t) + E_{\mathrm{pot}}(t) $, with $ E_{\mathrm{kin}} = \frac{1}{2} (m_{\mathrm{Li}} \dot{x}_{\mathrm{Li}}^2 + m_{\mathrm{Cr}} \dot{x}_{\mathrm{Cr}}^2 ) $, and $ E_{\mathrm{pot}} = \frac{1}{2} (m_{\mathrm{Li}} \omega_{\mathrm{Li}}^2 x_{\mathrm{Li}}^2 + m_{\mathrm{Cr}} \omega_{\mathrm{Cr}}^2 x_{\mathrm{Cr}}^2) $, respectively. As shown in the example in Fig.~\ref{fig:transient}, the evolution of $E_\mathrm{tot}(t)$ (black line) is in fact less affected by the oscillatory dynamics, such that a damping rate can be more easily extracted.
\begin{figure}[t]
    \centering
    \includegraphics[width=\linewidth]{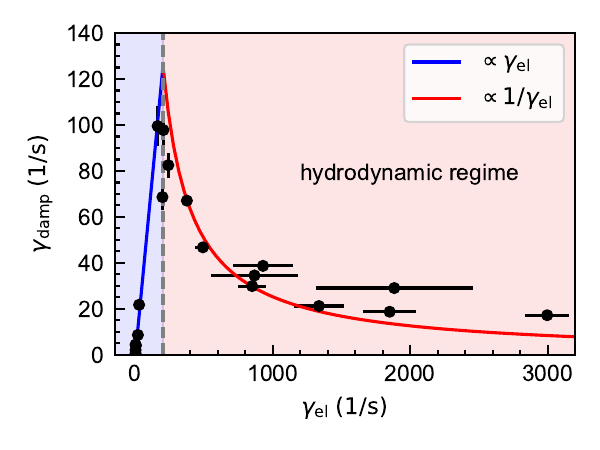}
    \vskip -20pt
    \caption{Identification of the transition from the collisionless to the collisionally-hydrodynamic regime via the analysis of the damping rate $\gamma_{\mathrm{damp}}$ as a function of $\gamma_\mathrm{el}$: A scaling of $\gamma_{\mathrm{damp}}\propto \gamma_\mathrm{el}$ is visible in the collisionless regime (blue line and shaded region), while the scaling $\propto 1/\gamma_\mathrm{el}$ characterizes the hydrodynamic regime (red line and shaded region). 
    Both lines are guides to the eye. 
    }
    \label{fig_energy_damping}
\end{figure}
Indeed, after an initial, short transient time of about 30\,ms, excluded from the fit, $E_\mathrm{tot}(t)$ exhibits a clean decay which is fitted by an exponential decaying function (red dashed line), from which we obtain $\gamma_\mathrm{damp}$ at all detunings. In Fig.~\ref{fig_energy_damping} we present the extracted damping rate $\gamma_\mathrm{damp}$, plotted as a function of the experimentally-determined  $\gamma_\mathrm{el}$. The observed trend is indeed found to follow the expected transition from collisionless to collisional hydrodynamics scaling \cite{PhysRevA.60.4734,PhysRevLett.87.173201,PhysRevLett.89.053202,FFerlaino_2003}, approximately once $\gamma_\mathrm{el}$ starts exceeding the axial trapping frequencies of the two atomic components. 
For our experimental conditions, this happens for magnetic-field detunings $\delta B \leq$ 30\,mG, below  which, as shown by the data presented in Fig.~\ref{fig_osc_together}, a rapid phase-locking of the two COMs dynamics is observed and, correspondingly,
$\gamma_{damp}$ is strongly reduced. We emphasize that this observation indicates our experimental capability to access the hydrodynamic regime not only along the weak, axial direction of our trap but, once  $\gamma_\mathrm{el} > \omega_{y,z} $, also along the transverse ones.
 
\section{Three-body rate coefficient and FR pole}
\label{Appendix:K3coeff&FRpoles}

In order to further characterize the stability of the atomic mixture in the strongly-interacting region, as well as to pinpoint the (absolute) magnetic-field location $B_0$ of the FR pole, we have performed additional studies of inelastic three-body recombination processes, besides those described in Sec.~\ref{Sec:FermiMixAndProcedures}. 
While such a survey has been conducted for each of the four high-field FRs occurring in all   Li$|i\rangle$-Cr$|j\rangle$ combinations with $i, j =$1, 2 \cite{PhysRevLett.129.093402}, in the following we summarize our experimental procedures and  findings by focusing on the absolute ground-state mixture  Li$|1\rangle$-Cr$|1\rangle$, and on the associated FR at 1414\,G. 
Conceptually, this additional characterization is based on protocols and analysis closely following those  already discussed in Sec.~\ref{Sec:FermiMixAndProcedures} to extract $\gamma_{loss}$: After preparing a weakly interacting mixture at a large and positive detuning from the FR pole, we quickly ramp the magnetic field towards the resonance region at small, variable $\delta B$ values, at which we monitor the subsequent atom number decay $N_{Cr}(t)$ of the Cr minority component as a function of time, evolving according with $\dot{N}_{\textrm{Cr}}=-K_3 \langle n^2_{\textrm{Li}} \rangle N_{\textrm{Cr}}=-\gamma_{loss}N_{\textrm{Cr}}$. 
Each magnetic field bias investigated is calibrated against the $\textrm{Li}|1\rangle\leftrightarrow\textrm{Li}|2\rangle$ RF transition, both before and after each decay measurement.

Contrarily to Sec.~\ref{Sec:FermiMixAndProcedures}, however, here we opt to investigate steady mixtures -- i.e. not exhibiting in-trap sloshing dynamics  -- that are prepared within our BODT trap, rather than in the sole IR beam.
On the one hand, this allows, at each magnetic field value, for a more accurate determination of the mean squared Li density $\langle n^2_{\textrm{Li}} \rangle_{\textrm{Cr}}$, now constant over time. 
This, combined with the $\gamma_{loss}$ values extracted from exponential fits to the Cr atom number evolution, as for the Sec.~\ref{Sec:FermiMixAndProcedures} measurement,  allows us to obtain, at each bias field, the three-body rate coefficient for the dominant Li-Li-Cr recombination processes as  $K_3=\gamma_{loss}/\langle n^2_{\textrm{Li}} \rangle_{\textrm{Cr}}$. 
On the other hand, exploitation of the green beam of our BODT, on top of the IR one, allows for an alternative, precise detection of the FR pole location $B_0$: Indeed, even a relatively low power level of the green light is found to induce strong photo-excitation losses once LiCr Feshbach dimers are formed, see Appendix \ref{Appendix:TrapLightLosses}. As such, once the magnetic field is lowered below $B_0$ on the molecular side of the FR, this light-induced, strong decay channel is opened, and it greatly overcomes the three-body collisional one. When this happens, the sample lifetime is markedly reduced, hence yielding an enhanced $\gamma_{loss}$. 
This experimental protocol thus allows us to accurately determine both the three-body recombination rate coefficient $K_3$ for $B\geq B_0$, as well as to obtain an additional measure of the FR pole $B_0$ -- besides the one obtained from the fit of $\gamma_{el}$ presented in Sec. \ref{Sec:FermiMixAndProcedures} -- solely based on the study of inelastic losses. 

\begin{figure}[t]
    \centering
    \includegraphics[width=\columnwidth]{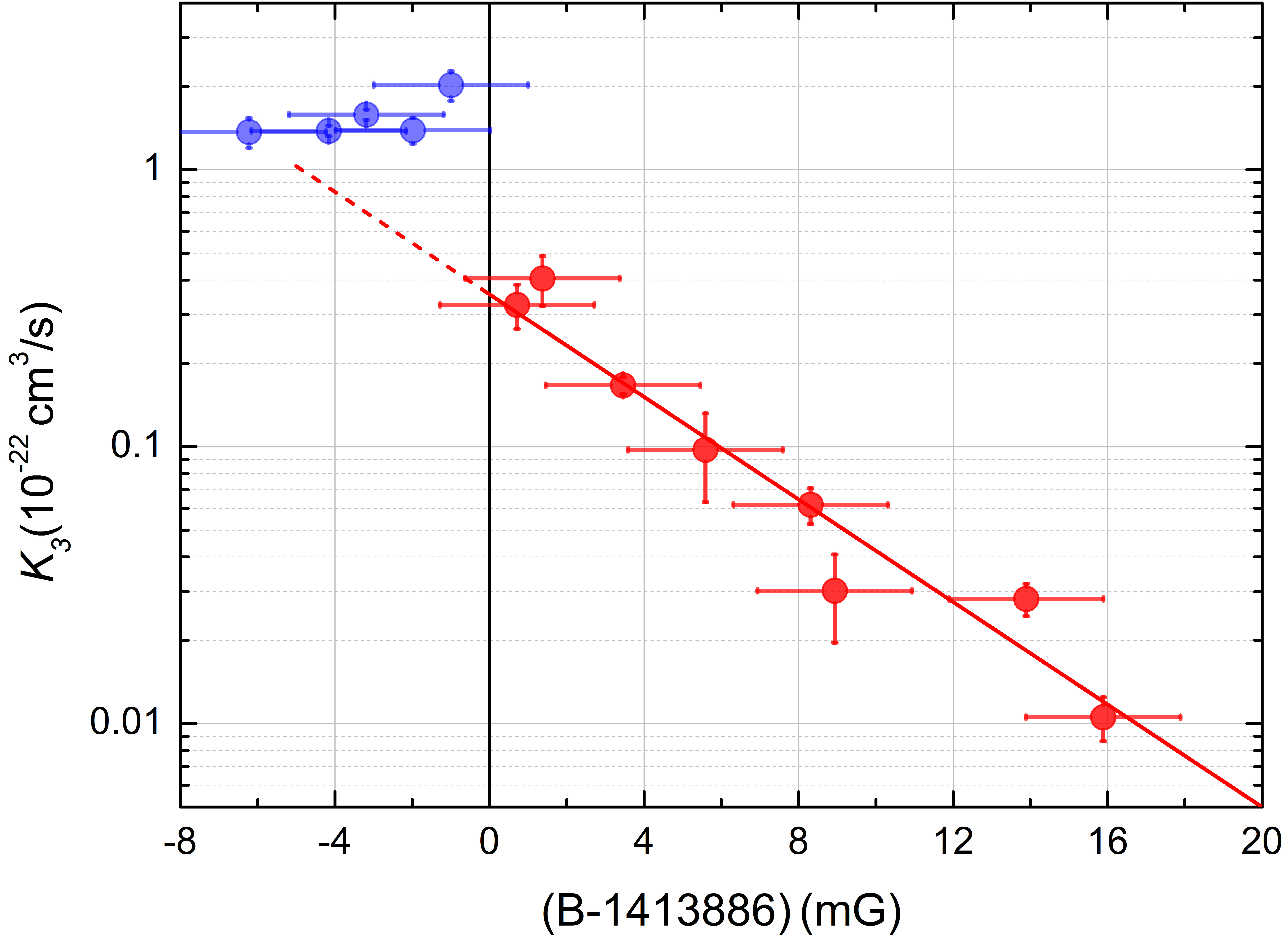}
    \caption{Three-body recombination rate coefficient for Li-Li-Cr collisions as a function of magnetic field across the FR region, obtained for mixtures prepared in the BODT trap by following the analysis described in the text. Strong photo-excitation losses, induced by the green light of our trap for $\delta B\!\leq\!0$, results in a sharp increase of the derived $K_3$, marked by the different color of the experimental data. This enables us to pinpoint the absolute resonance pole position $B_0$ with a 5\,mG accuracy, see text and Table \ref{tbl:ResLocations}. Red line is the best-fit of the data above the FR pole to Eq.~(\ref{K3BCS}), fixing $\delta \tilde{\mu}=0.77 \times 2 \mu_\textrm{B}$ while leaving  both $T$ and $K_3^0$ as free parameters.}
    \label{fgr:K3COeff}
\end{figure}

The results of this characterization are summarized in Fig.~\ref{fgr:K3COeff}, where we show the experimentally-determined $K_3$ coefficient as a function of the magnetic field across the resonance region. 
We interpret the sudden jump between the red and blue data points as the crossing of the resonance pole, below which Feshbach dimers are formed and quickly lost due to the fast trap-induced photo-excitation process, see  Appendix \ref{Appendix:TrapLightLosses}. 
The jump is centered at $B_0$=1413.886(5)\,G, and the error budget accounts for residual AC field noise, calibration of the COMP coils, magnetic-field inhomogeneity and residual long term drifts, in order of importance.
Experimental data marked in blue cannot be anymore interpreted as a $K_3$ coefficient for three-body collisions and their interpretation goes beyond the scope of our work. 
At positive detunings (i.e. for $B>B_0$), and similarly to the  $\gamma_{loss}$ trend presented in Sec.~\ref{Sec:FermiMixAndProcedures}, the extracted $K_3$ exhibits an exponential growth as the resonance pole is approached from above, $\delta B \rightarrow 0^+$, in qualitative agreement with previous observation on homonuclear Fermi mixtures near narrow Feshbach resonances \cite{PhysRevLett.120.193402}.

Following the theoretical analysis of Ref. \cite{PhysRevLett.120.193402}, valid for an infinitely narrow FR ($R^* \rightarrow \infty$), the $K_3$ coefficient for a thermal mixture is expected to vary, for $\delta B\!\geq\!0$, as:
\begin{equation}
\label{K3BCS}
K_3(\delta B, T)=K^0_3(T)\, \exp\left(- \frac{\delta \mu \delta B}{k_B T}\right).
\end{equation}
Here $K^0_3(T)$ is the (temperature-dependent) maximum value of $K_3$ reached at the resonance pole, $\delta \mu$ is the differential magnetic moment between the open and close channels associated with the FR,  $T$ is the gas temperature and $k_B$ the Boltzmann constant. 
In case of degenerate samples, one can still employ Eq.~(\ref{K3BCS}) by defining an effective temperature $T_{eff}$ that connects with the mean kinetic energy of the system. In our case, this is essentially set by the majority Li Fermi gas, such that we evaluate $T_{eff}$ by setting $3/2 k_B T_{eff}=\langle E_{kin,Li}(T/T_F) \rangle$. Taking the experimentally-determined value of $T/T_F$= 0.27(2), with $T_F$= 1.25(5)\textmu K, and considering that for an ideal Fermi gas $\langle E_{kin,Li}(0.27) \rangle\sim 0.55 k_B T_F$, we obtain $T_{\textrm{eff}}\sim$460\,nK.

However, by fixing $T_{\textrm{eff}}$ to this value and setting the magnetic moment to $\delta \mu = 2 \mu_\textrm{B}$, i.e to the value characterizing each of the high-field FRs of Li-Cr \cite{PhysRevLett.129.093402}, Eq.~(\ref{K3BCS}) yields a trend significantly steeper than the one experimentally-determined. Interestingly, this mismatch arises from the fact that our resonance is indeed not infinitely narrow, featuring a sizable though finite $R^*$ value. Extension of Ref. \cite{PhysRevLett.120.193402} theory to finite effective-range values (D. Petrov, unpublished) yields a  trend for $K_3$ qualitatively analogous to that of Eq.~(\ref{K3BCS}), but with an effective magnetic moment $\delta \tilde{\mu}$ that is progressively decreased for decreasing $R^*$ values. In particular, such an extended theory model predicts for the Li-Cr FRs here investigated, featuring $R^*\sim$6000 $a_0$, a $\delta \tilde{\mu}=0.77 \delta \mu$. 
A functional fit of Eq.~(\ref{K3BCS}) to the red dataset in Fig.~\ref{fgr:K3COeff}, where we set $\delta \tilde{\mu}=0.77 \times 2 \mu_\textrm{B}$ while leaving both $T$ and $K^0_3$ as free parameters, is plotted in the figure as a solid red line. The fit, that reproduces our data remarkably well, returns $K^0_3=0.35 (5) \times 10^{-22}\,$cm$^6$\,Hz and $T=490(40)\,$nK, the latter matching within uncertainty the estimated $T_{\textrm{eff}}$ value.
A fit based on the same extended theory model is employed to analyze the $\gamma_{loss}$ data shown in Fig. ~\ref{fig_hydro}(c) of Sec.~\ref{Sec:FermiMixAndProcedures}. Similar protocols were employed to pinpoint the pole of all four FRs occurring between the two lowest spin states of Li and Cr at high fields, and the results of this characterization are summarized in Table \ref{tbl:ResLocations}.
\begin{table}[t]
\small
\begin{tabular*}{\columnwidth}{@{\extracolsep{\fill}}lll}
\hline
\noalign{\smallskip}
 & Cr$|1\rangle$ & Cr$|2\rangle$\\
\hline
\noalign{\smallskip}
Li$|1\rangle$ & 1413.886(5) & 1417.937(30)\\
Li$|2\rangle$ & 1460.933(5) & 1464.159(30)\\
\hline
\end{tabular*}
\caption{Precise determination, through loss-spectroscopy measures in the BODT, of the magnetic-field location of the high-field $s$-wave Feshbach  resonances  for all   Li$|i\rangle$-Cr$|j\rangle$ combinations with $i, j =$1, 2, already identified in Ref. \cite{PhysRevLett.129.093402} with lower accuracy. Error budget,  in brackets, accounts for residual AC field noise, calibration of the COMP coils, magnetic-field inhomogeneity and residual long term drifts, in order of importance.}
\label{tbl:ResLocations}
\end{table}

\begin{figure}[t]
\centering
  \includegraphics[width=\columnwidth]{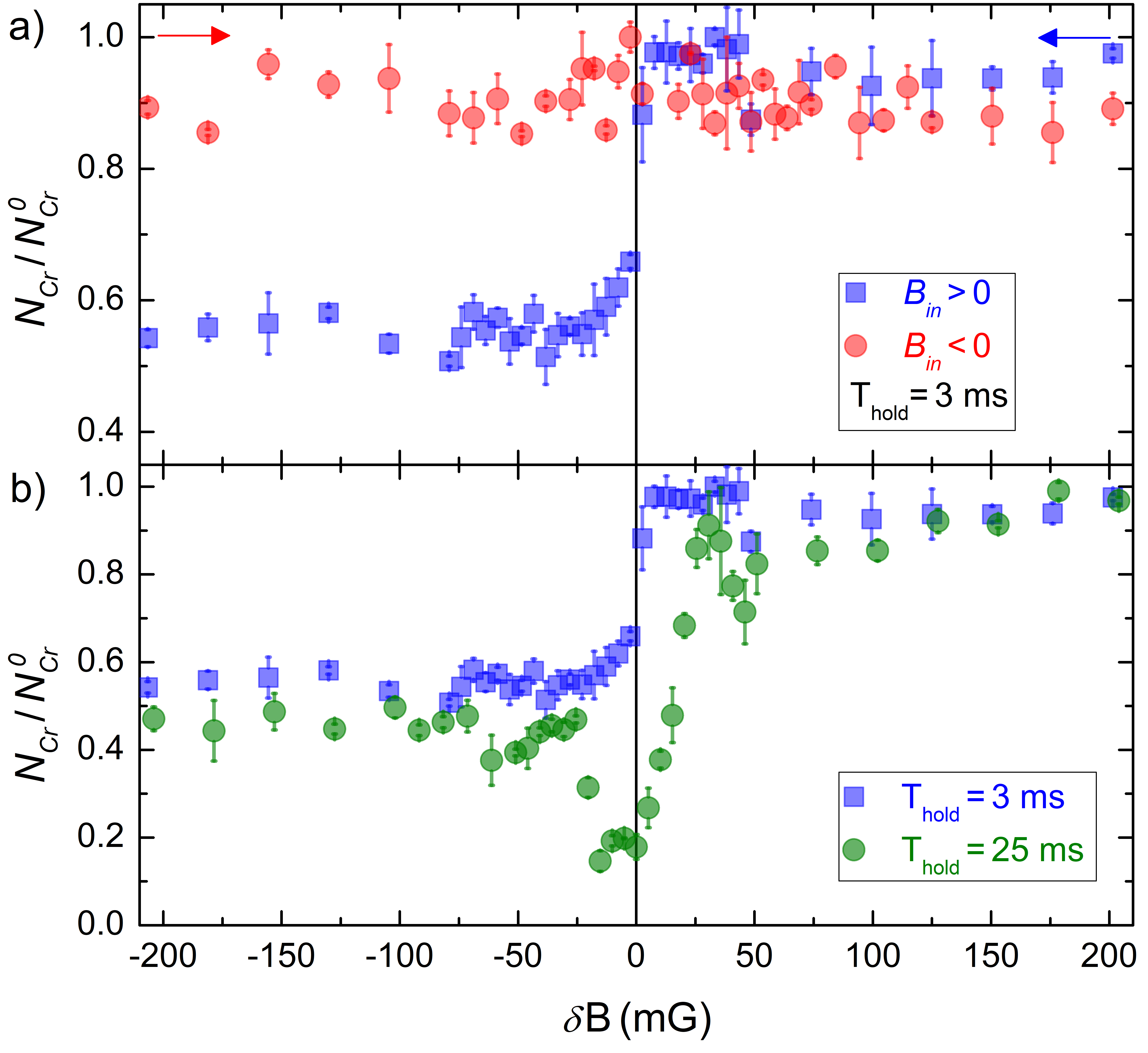}
  \caption{(a) Fractional Cr loss  measured in the BODT as a function of the magnetic field detuning, that is reached through a ramp starting at a bias field $B_{\mathrm{in}}$ either far above (blue squares), or far below (red circles) the FR pole. The signal is detected after a hold time at the final field of 3\,ms. (b) Comparison of the Cr loss line-shapes, obtained with magnetic-field ramps starting from $B_{\mathrm{in}}>$0, and measured after a hold time $T_{\mathrm{hold}}\!=\!3\,$ms (blue squares) and $T_{\mathrm{hold}}\!=\!25\,$ms (green circles).}
  \label{fgr:LineshapeSummary}
\end{figure}

Finally, we remark how the strong photo-excitation loss induced by the green BODT beam represents a valuable tool to monitor molecule formation through negative signals at short hold times $T_{\mathrm{hold}}$, unaffected by spurious offset induced by three-body loss processes. This is demonstrated by the data presented in Fig.~\ref{fgr:LineshapeSummary}(a). There, we show the Cr loss signals measured at $T_{\mathrm{hold}}$= 3 ms as a function of the final detuning, which is reached through magnetic field ramps starting either far above ($B_{\mathrm{in}}\!>\!0$, blue squares) or far below ($B_{\mathrm{in}}\!<\!0$, red dots) the FR pole. In the former configuration -- that allows for magneto-association of LiCr dimers -- we reveal a strong drop of the Cr signal as the FR pole is crossed. In contrast, in the latter case -- for which molecule formation can only occur via three-body processes, and only collisional losses may reduce the signal -- the Cr population remains constant within experimental noise throughout the resonance region. It is thus legitimate to attribute the observed drop of the blue data solely to molecule formation via magneto-association, since spurious effects due to three-body processes are negligible at short $T_{\mathrm{hold}}$ values. 
Indeed, collisional losses, which add to those induced by photo-excitation of LiCr dimers, become sizable only at significantly longer hold times. This is testified by  Fig.~\ref{fgr:LineshapeSummary}(b), where the magneto-association line-shape obtained with $T_{\mathrm{hold}}\!=\!3\,$ms (blue squares), and shown already in Fig.~\ref{fgr:LineshapeSummary}(a), is compared with the one measured after a hold time of 25\,ms (green circles). As one can see, only for such long $T_{\mathrm{hold}}$ values three-body losses start yielding,  near the resonance pole, a signal contribution comparable with the one due to molecule formation.

\section{Probing Feshbach dimer properties through absorption imaging on atomic transitions}
\label{Appendix:AbsImgFBmols}

In the following we provide further details on the model introduced in Sec.~\ref{Sec:MoleculeProperties_OpenCloseFractions}, and on the data analysis that allows us to determine both open-channel fraction and binding energy of LiCr dimers through absorption images acquired with laser light resonant with atomic transitions. As explained in Sec.~\ref{Sec:MoleculeProperties_OpenCloseFractions}, we assume that the atomic light interacts with Feshbach dimers only via a non-negligible open-channel fraction $1-Z$ (assumption (i)), that the dimer dissociation rate follows the Fermi's golden rule $\gamma_b=\gamma_a (1-Z)$ (assumption (ii)), and, finally, that the imaging is performed on a cycling transition with low saturation parameter $s_a \ll 1$ (assumption (iii)). In this case the instantaneous optical density of the initially pure molecular sample can be written, see Eq.~\ref{Eq:TimeDependentOD} of Sec.~\ref{Sec:MoleculeProperties_OpenCloseFractions}, as:
\begin{equation*}
OD(t)= OD_a (1-e^{-\gamma_b t}).
\end{equation*}
Here $OD_a$ is the optical density of an atomic sample of equivalent column density $OD_a=\sigma_a n^0_{2D,b}$, with $\sigma_a$ being the field-dependent absorption cross section and $n^0_{2D,b}$ the molecule column density.

We now derive the observed optical density after an absorption imaging pulse of duration $\tau_p$ from the instantaneous time-dependent optical density $OD(t)$ above. We recall that the temporal variation of photon counts $N_{p}$ on the camera during the probe pulse shining onto the Feshbach molecule sample follows the relation
\begin{equation}
\label{PhotonCountsDer}
\partial_t N_{p,b}= C \times I_0  \times e^{-OD (t)},
\end{equation}
where $C$ is the overall quantum efficiency of the imaging system and $I_0$ is the laser intensity. At the end of the pulse, then, we have
\begin{equation}
\label{PhotonCounts}
N_{p,b} (\tau_p) = C I_0 e^{-OD_a} \int_0^{\tau_p} e^{OD_a e^{-\gamma_b t}} dt,
\end{equation}
in contrast with the case of a pure atomic sample, for which $N_{p,a} (\tau_p) = C I_0 \tau_p e^{-OD_a}$.
In both cases, the reference image, with neither dimers nor atoms, yields $N^0_{p,b}=C I_0 \tau_p$ counts. Using the definition of optical density $OD=-\ln(N_p/N^0_p)$, taking as reference the atomic optical density at resonance $OD_a^0$, we derive the following suppression factor
\begin{equation}
\label{ODSuppressionMolGeneral}
\frac{\overline{OD}(\tau_p)}{OD^0_a}=\frac{\gamma_a}{\gamma^0_a}\times \left( 1-\frac{1}{OD_a} \ln \left( {\frac{1}{\tau_p} \int_0^{\tau_p} e^{OD_a e^{-\gamma_b t}} dt} \right) \right),
\end{equation}
\begin{equation}
\label{ODSuppressionMolGeneralgamma0}
\gamma^0_a=\gamma(s_a,\Gamma_a,0),
\end{equation}
\begin{equation}
\label{ODSuppressionMolGeneralgammab}
\gamma_b=\gamma_a(s_a,\Gamma_a,\delta_a+\epsilon_b/h) \times (1-Z),
\end{equation}
where the $B$-field dependence of $Z$, $\delta_a$ and $\epsilon_b$ is implicit. In our case we have $\delta_a(B)=\mu_\textrm{B} \delta B$ for the Zeeman shifts of the imaging transitions of both Li and Cr. In the limit of low optical density, $OD^0_a\!\ll\!1$, Eq.~(\ref{ODSuppressionMolGeneral}) simplifies to Eq.~(\ref{eq:ODSuppressionMolLowOD}) of Sec.~\ref{Sec:MoleculeProperties_OpenCloseFractions}. 

Most importantly, our model can be exploited to extract the open-channel fraction at any detuning, without \emph{a-priori} knowledge of the functional form of $Z(\delta B)$ and $\epsilon_b(\delta B)$. As long as the binding energy of the Feshbach molecule is negligible with respect to the linewidth, $\epsilon_b/h \ll \Gamma_a$, Eq.~(\ref{ODSuppressionMolGeneral}) only has $Z(\delta B)$ as unknown quantity and can easily be inverted $N_i(\delta B_i) \rightarrow Z_i(\delta B_i)$. If this condition does not hold, we are left, at each $\delta B_i$ value, with two unknown quantities $Z_i$ and $\epsilon_{b,i}$ which cannot be uniquely inferred from $N_i$. To circumvent this limitation, and without loss of generality, we make use of the properties of the Feshbach state summarized by the Hellman-Feynman theorem Eq.~(\ref{eq:HFtheorem}) and the limiting behaviour for $\delta B\rightarrow0$ of the magnetic moment and binding energy, $\delta \mu_b\rightarrow0$ and $\epsilon_b\rightarrow0$, respectively. Thus, instead of extracting from each measurement point $N_i(\delta B_i) \rightarrow Z_i(\delta B_i)$, we iteratively run over consecutive measurement points starting from the largest positive detuning $\max (\delta B_i)$ and moving down to $\min (\delta B_i)$. The algorithm estimates the binding energy at $\delta B_i$ as $\epsilon_{b,i}=\epsilon_{b,i-1}+Z_{i-1} \delta \mu (\delta B_{i-1}-\delta B_i)$ and extracts a new open-channel fraction $\{ N_i,\epsilon_{b,i} \}\rightarrow Z_i$. The algorithm is initialized with $(Z_{i0}=0,\epsilon_{b,i0}=0)$ and forces $(Z_{i}=0,\epsilon_{b,i}=0)\quad\forall\quad\delta B_i>0$.

\section{Trap-light-induced losses}
\label{Appendix:TrapLightLosses}
\begin{figure}[t]
\centering
\includegraphics[width=\columnwidth]{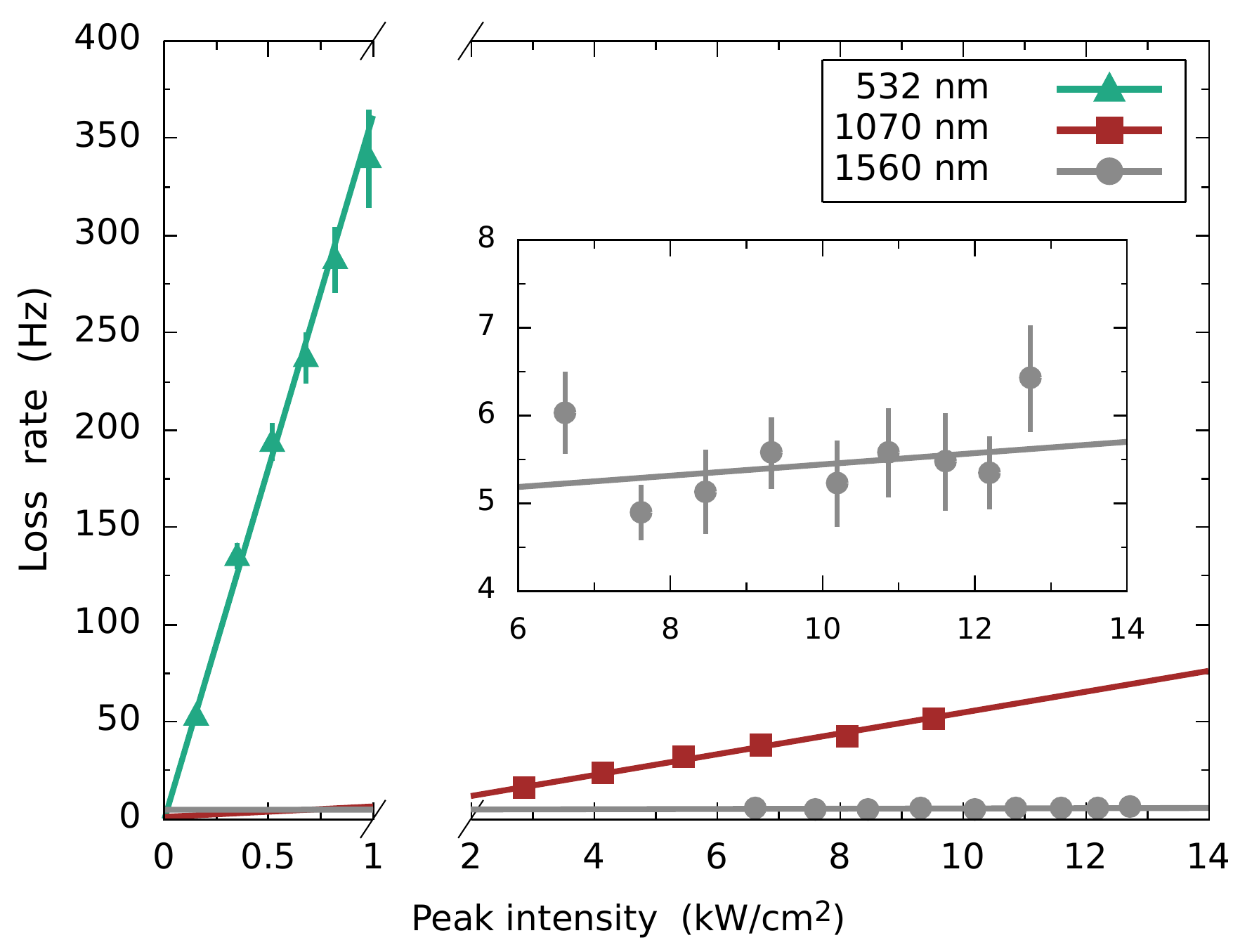}
 \caption{Loss rates as function of intensity for the trap wavelengths explored in this work. Green triangles, red squares, and gray circles show the experimental exponential loss rates for 532\,nm, 1070\,nm, and 1560\,nm, respectively. Best linear fits to the data are plotted as solid lines with matching color code.}
 \label{fgr:ODT_llossrates}
\end{figure}
The BODT configuration exploited in Sec. \ref{Sec:MoleculeFormation} to reach optimal magneto-association efficiency and high PSD of LiCr molecules strongly constraints the molecule lifetime in the few ms time scale and is not suitable for trapping LiCr Feshbach molecules. In order to characterize this effect and decouple it from the simultaneous collisional losses, we purify the molecule sample right after molecule formation by removing the leftover atoms. Li$|1\rangle$ is removed via RF pulse $\textrm{Li}|1\rangle \rightarrow \textrm{Li}|2\rangle$ plus optical blast, while Cr$|1\rangle$ is spilled from the trap, which for this species is significantly shallower than for both Li and LiCr components \cite{MolPolarizabilities}.
 
We first isolate the effect of the multimode-1070\,nm ODT beam by ramping down the green ODT power during molecule association. We then ramp the IR power to a variable value, where we record the subsequent molecule number drop. Exponential fits to the decay data return the loss rates, which are reported as a function of the IR laser intensity in Fig.~\ref{fgr:ODT_llossrates} (red squares) together with a linear fit to the data. Using the fitted slope and taking into account a finite open-channel fraction $1-Z>0$ at the probe detuning $\delta B=-100\,$mG according to Ref.~\cite{PhysRevResearch.5.033117}, we derive $\Gamma_{CC}=5.9(2)\,\textrm{Hz}/(\textrm{kW}\,\textrm{cm}^{-2})$. A similar analysis carried out on molecular samples realized in the 1560\,nm beam yields instead a photoexcitation rate consistent with zero within experimental uncertainty, see gray circles in Fig.~\ref{fgr:ODT_llossrates} and corresponding inset. 

Since the 532\,nm green beam is anti-trapping for our molecules, we study its effect by ramping it up to a variable power while keeping the IR ODT power fixed. The corresponding experimental loss rates are shown in Fig.~\ref{fgr:ODT_llossrates} as green triangles and yield a significantly higher slope of $\Gamma_{CC}=397(14)\,\textrm{Hz}/(\textrm{kW}\,\textrm{cm}^{-2})$. This value, about two orders of magnitude larger than the IR one, confirms the detrimental effect of 532\,nm light on Feshbach molecules and the strong constraint it puts on the possible timescale for manipulation of the LiCr sample in the combined BODT. We note that, for each dataset and trapping light,  the systematic error of $\Gamma_{CC}$,  arising from intra-species LiCr-LiCr 
inelastic collisions, is negligible because the dimer density is kept fixed at all power levels explored within our experimental accuracy.

In conclusion of this section, we also remark that the measured dependence of the one-body loss rate upon the trap-light wavelength points to a stronger off-resonant photon scattering for increasing photon energy. Our findings qualitatively agree with our \emph{ab initio} calculations, presented in Sec.~\ref{Sec:Theory}, that find a progressively higher density of molecular states, coupled via electric-dipole transitions to the Feshbach state, for increasing photon energy.


\begin{thebibliography}{92}
\bibitem [{Ato()}]{AtomicPolarizabilities}%
  \BibitemOpen
  \href@noop {} {}\bibinfo {note} {The dynamic polarizabilities relevant for
  this work are $\alpha_{\textrm{Li}}(532\,\textrm{nm})=-273\,$a.u,
  $\alpha_{\textrm{Li}}(1070\,\textrm{nm})=270\,$a.u.,
  $\alpha_{\textrm{Li}}(1560\,\textrm{nm})=201\,$a.u. for Li, and
  $\alpha_{\textrm{Cr}}(532\,\textrm{nm})=164\,$a.u.,
  $\alpha_{\textrm{Cr}}(1070\,\textrm{nm})=89\,$a.u.,
  $\alpha_{\textrm{Cr}}(1560\,\textrm{nm})=83\,$a.u. for Cr.}\BibitemShut
  {Stop}%
\bibitem [{Mol()}]{MolPolarizabilities}%
  \BibitemOpen
  \href@noop {} {}\bibinfo {note} {Dynamic polarizabilities of Feshbach LiCr
  are expected to be the sum of the atomic ones:
  $\alpha_{\textrm{LiCr}}(532\,\textrm{nm})=-109\,$a.u,
  $\alpha_{\textrm{LiCr}}(1070\,\textrm{nm})=359\,$a.u.,
  $\alpha_{\textrm{LiCr}}(1560\,\textrm{nm})=284\,$a.u., of which the last two
  are consistent with our measurements.}\BibitemShut {Stop}%
\end{thebibliography}

%

\end{document}